\documentclass[aps,twocolumn,superscriptaddress,prb,floatfix,citeautoscript,preprintnumbers]{revtex4-1}
\usepackage{graphicx,color}
\usepackage{amssymb}
\usepackage{amsmath}
\usepackage{epstopdf,epsfig}
\usepackage[normalem]{ulem}
\usepackage{comment}
\usepackage{breakcites}
\usepackage[breaklinks=true]{hyperref}
\usepackage[caption=false]{subfig}
\usepackage{enumitem}

\usepackage{bbold,dsfont}
\def\Id{\mathbb{1}}
\def\His{H_{\mathrm{I}}}
\def\Hxyz{H_{\mathrm{XYZ}}}
\def\Z2{Z_{\mathrm{st2}}}
\def\Dtr{D_{\mathrm{tr}}}

\newcommand{\beq}{\begin{equation}}
\newcommand{\eeq}{\end{equation}}
\newcommand{\bal}{\begin{align}}
\newcommand{\eal}{\end{align}}
\newcommand{\ra}{{\rightarrow}}

\newcommand{\nn}{{\nonumber}}
\def\bigO{{O}}

\newcommand{\ket}[1]{\mbox{$ | #1 \rangle $}}
\newcommand{\bra}[1]{\mbox{$ \langle #1 | $}}
\newcommand{\ev}[1]{\langle #1 \rangle}

\begin{document}

\title{How much entanglement is needed to reduce the energy variance?}
\begin{abstract}
We explore the relation between the entanglement of a pure state and its energy variance
for a local one dimensional Hamiltonian, as the system size increases.
In particular, we introduce a construction which creates a matrix product state of arbitrarily small energy variance
$\delta^2$ for $N$ spins, with bond dimension scaling as $\sqrt{N} D_0^{1/\delta}$, where
$D_0>1$ is a constant. This implies that a polynomially increasing bond dimension is enough to construct
states with energy variance that vanishes with the inverse of the logarithm of the system size.
We run numerical simulations to probe the construction on two different models,
and compare the local reduced density matrices of the resulting states to the corresponding thermal
equilibrium.
Our results suggest that the spatially homogeneous states with logarithmically decreasing variance, which can be constructed efficiently,
do converge to the thermal equilibrium in the thermodynamic limit,
while the same is not true if the variance remains constant.
\end{abstract}

\author{Mari Carmen Ba\~nuls}
\affiliation{Max-Planck-Institut f\"ur Quantenoptik, Hans-Kopfermann-Str.\ 1, D-85748 Garching, Germany}
\affiliation{Munich Center for Quantum Science and Technology (MCQST), Schellingstr. 4, D-80799 M\"unchen}
\author{David A. Huse}
\affiliation{Physics Department, Princeton University, Princeton, New Jersey 08544, USA}
\author{J. Ignacio Cirac}
\affiliation{Max-Planck-Institut f\"ur Quantenoptik, Hans-Kopfermann-Str.\ 1, D-85748 Garching, Germany}
\affiliation{Munich Center for Quantum Science and Technology (MCQST), Schellingstr. 4, D-80799 M\"unchen}


\maketitle

\section{Introduction}
\label{sec:intro}

Entanglement plays a central role in several phenomena in many-body quantum systems.
Ground and low excitation states of local lattice Hamiltonians typically have very little entanglement as they fulfill an area law\cite{Hastings2007,eisert2010area}: the entanglement of a connected region with the rest scales with the number of particles (area) at its boundary.
As a consequence, those states can be efficiently described with tensor networks, with a number of parameters that only scales polynomially with the total system size.\cite{verstraete2006faithfully,Hastings2006a,hastings07entropy}
However, generic eigenstates may possess a lot of entanglement, as they are expected to obey a volume law.
This has relevant implications,
like the fact that the dynamics of quenched systems is hard to describe in terms of tensor networks, at least for sufficiently long times\cite{Osborne2006,Schuch2008a}.
This volume law is also closely related to the eigenstate thermalization hypothesis\cite{deutsch91eth,srednicki94eth,alessio2016eth,deutsch2018review}, namely the fact that generic eigenstates are able to capture the local properties of systems in thermal equilibrium, when the number of lattice sites $N\to\infty$: the entropy of any finite region in the thermodynamic limit must be extensive, and thus entanglement has to obey a volume law.  \cite{gemmer2001second,popescu2006nphys,kaufman2016}
Indeed, for a large variety of local Hamiltonians it is expected that local expectation values in any eigenstate with energy in the interval $\Delta_E=[E- a N^\alpha, E+ a N^\alpha]$  converge to the same value in the thermodynamic limit for any constant $a$ and $\alpha<1$.

The observation that all eigenstates in an energy interval lead to the same properties in the thermodynamic limit naturally implies that {\em convex} combinations thereof fulfill the same property.
Those mixed states may have an energy variance $\delta^2$ that scales according to $\delta\sim N^\alpha$ and still give rise to thermal averages.
However, this is not necessarily the case for {\em linear} combinations of those eigenstates or, more generally, for arbitrary pure states with $\delta\sim N^\alpha$.
In fact, product pure states typically possess a variance that scales as $\delta \sim N^{1/2}$ and do not have any entanglement at all, so that they cannot describe thermal properties of a system.
This raises some natural questions: Are there states with $\delta$ below $\sqrt{N}$ but still with little entanglement? Is there a general relation between those two quantities?
How does the energy variance of a pure state need to scale with $N$ in order to ensure that the state describes local thermal properties?

Some related problems have been studied in the literature.
 Typicality arguments\cite{popescu2006nphys} can be invoked to
argue that most pure states compatible with any macroscopic constraints will exhibit very similar local expectation values.
Under appropriate considerations, this is enough to ensure that energy eigenstates look locally thermal\cite{goldstein2006cantyp}.
However, if the most strict sense of typicality is considered\cite{hamazaki2018atyp,dymarsky2019canuni}, this is, if all eigenstates within an energy shell need to have expectation values that are exponentially close (in the system size) to
the shell average, this requires exponentially small width of the energy shell for most few-body Hamiltonians\cite{hamazaki2018atyp}.
For chaotic systems, a polynomial relation was found\cite{dymarsky2019canuni} between the energy width and the maximal deviation from thermal behavior for any state supported on it, in particular, in terms of the
subsystem entropy.
Additionally, the typical entanglement of energy eigenstates
has been the focus of recent theoretical studies for
chaotic and integrable systems\cite{vidmar2017entropy,vidmar2017fermions,lu2019renyi,huang2019eigen,hackl2019avrg}.
Less is known about the typical behavior of entanglement for a
pure state that is a superposition of energy eigenstates within a narrow energy shell,
although results exist that characterize the typical entropy of physical states, understood as those that can be reached by unitary evolution with local Hamiltonians \cite{poulin2011illusion,hamma2012physical,hamma2012pra}.

A related topic is
the concept of \emph{thermal pure quantum states}\cite{sugiura12pure,sugiura13canonical,hyuga2014mb}  (TPQ), introduced
to develop a pure state formulation of statistical mechanics.
TPQ are random states for which expectation values of local observables 
probabilistically converge, in the thermodynamic limit, to their values in a given statistical equilibrium ensemble.
The variance of their energy density distribution vanishes as $1/N$.
In\cite{garnerone2013pureTD} TPQ states
were constructed starting with a state drawn from a random matrix product state (RMPS) ensemble\cite{garnerone2013micromps} with fixed bond dimension,
what allows for a more efficient sampling in many-body systems.
However, if the resulting TPQ  needs to approach the entanglement content (for a subsystem) that characterizes
the equilibrium ensemble, the accurate MPS approximation of the TPQ will be unfeasible in most cases.

In this paper we address the question of how much entanglement is required to reduce the energy variance of a local Hamiltonian in one dimensional spin chains.
In particular, we construct states with an arbitrary value of the variance $\delta^2$ and with entanglement that scales as $(k/\delta)+\log \sqrt{N}$, where $k$ is a constant.
We use matrix product state (MPS) techniques\cite{verstraete08algo,schollwoeck11age}  for the deterministic construction and also to compute the entanglement, which we estimate through the bond dimension of the states.
We also extensively check numerically this prediction with the Ising model in a transverse and longitudinal field, and with the Heisenberg model.
Our results imply that it is indeed possible to build states of constant variance $\delta=\bigO(1)$ but with little entanglement, which grows only logarithmically with $N$ and a bond dimension of the MPS that grows polynomially. In fact, one can even take $\delta \simeq 1/\log(N)$ while keeping such polynomial scaling with $N$. However, if we want to obtain a scaling $\delta\propto 1/N$, the entanglement will grow linearly with the size. We also investigate to what extent the states we construct can recover the thermal properties as $N$ grows. Since the entropy in thermal equilibrium is extensive, a necessary condition for a region of size $L$ to be thermal is that its entanglement entropy is $\bigO(L)$. Thus, the bond dimension of the MPS must at least scale exponentially with $L$. If one restricts the bond dimension to grow only polynomially with $N$, the largest thermal region can be at most $L \sim O(\log N)$. 
According to our bounds, the required $\bigO(L)$ entropy can be thus achieved with a variance that decreases as $\delta\sim 1/\log(N)$.
Our numerical results confirm that  we can decrease the variance as $\delta \sim 1/\log(N)$ keeping a polynomially scaling bond dimension and,
for fixed value of $L$, all local observables in the region of size $L$ converge to their thermal values in the thermodynamical limit.

The rest of the paper is organized as follows. In section~\ref{sec:prelim} we introduce the general
aspects of our setup and our notation.
Then, section~ \ref{sec:construction} presents two explicit constructions of pure states
which can attain arbitrarily small energy variance.
In \ref{sec:bounds} we derive our main result: the bounds on bond dimension
and entropy of such states as a function of the variance.
We also discuss the limitations and implications of our results.
Section~\ref{sec:num} presents our numerical results for the Ising and Heisenberg spin chains.
Besides checking the scaling of the variance, and the verification of the theoretically derived bounds, we probe the convergence of reduced density matrices
towards the thermal equilibrium states.
In section~\ref{sec:variational} we compare the results with those from variational minimization of the variance for the
same sets of parameters.
Finally in section~\ref{sec:discussion} we discuss our results and summarize our conclusions.

\section{Preliminaries}
\label{sec:prelim}

We are interested in analyzing the entanglement required to achieve a given energy variance $\delta^2$.
We consider a spin chain of local dimension, $d$, and a local Hamiltonian
 \beq
 \label{H}
 H=\sum_{n=1}^{N} h_i
 \eeq
where $h_n$ acts on spins $n$ and $n+1$. We will consider open boundary conditions, i.e. $h_N$ acts only on the $N$-th spin, although our results can be easily extended to the case of periodic boundary conditions. Without loss of generality, we will assume that ${\rm tr}(h_n)=0$, and ${\rm tr}_n(h_n)=0$. Note that if the latter is not the case, we can always include the part that does not vanish in the term $h_{n+1}$. We will normalize the Hamiltonian such that
 \beq
 \max_n \|h_n\|=1,
 \eeq
where we took the operator norm, so that its spectrum, $\sigma(H) \subseteq[-N,N]$. When we consider sequences of Hamiltonians with increasing number of spins, will also assume that ${\rm min} \|h_n\|= h_{\rm min}>0$, where $h_{\rm min}$ is some constant independent of $N$. In particular, for the numerical computations we will take $h_n=h_m$ (except for $n=N$), so that this is automatically fulfilled.

The energy variance of a pure state $\Psi$ is
 \beq
 \delta^2 = {\bra{\Psi} H^2 \ket{\Psi}- \langle\Psi|H\ket{\Psi}^2}
 \eeq
In order to analyze the entanglement present in the state, one can consider a ``cut"
of the $\ell$-th link of the chain, with $\ell \in\{1,\ldots N-1\}$. This divides the chain in two regions, which we denote by the number of spins they contain, respectively $\ell$ and $N-\ell$. The entanglement entropy
with respect to this bipartition is
\beq
 S_{\ell}=- {\rm tr} (\rho_{\ell} \log_2\rho_{\ell}) = S_{N-\ell},
 \label{eq:SA}
 \eeq
 where the reduced state $\rho_{\ell}=\mathrm{tr}_{\ell+1,\ldots N} \ket{\Psi}\bra{\Psi}$.

This quantity is bounded by $\log_2 d$ times the minimum between the number of spins in both regions. Typically, when the sizes are large $S_{\ell}$ is difficult to compute (as it requires diagonalizing $\rho_{\ell}$), and even to bound (because two states that are arbitrarily close may have very different values). Alternatively, we will also consider the bond-dimension required to describe the state $\Psi$ in terms of a matrix product state (MPS). That is, the size, $D$, of the matrix $A^s[n]$ such that $\ket{\Psi}\approx \ket{\Phi_D}$ with
 \beq
 |\Phi_D\rangle = \sum_{s_1,\ldots,s_N=1}^d {\rm tr}\left( A^{s_1}[1] \ldots A^{s_1}[N]\right) |s_1,\ldots,s_N\rangle
 \eeq
Typically, one would expect that the maximum entropy with respect to all possible cuts, $S=\max_{\ell} S_{\ell}$, fulfills
 \beq
 \label{SA}
 S \sim c \log(D)
 \eeq
for some $c=\bigO(1)$.

In the following sections, we will construct states with an arbitrary energy variance $\delta^2$ and that  (for large system sizes and small $\delta$)
can be approximated by an MPS with bond dimension
 \beq
 \label{bonddim}
 D \lesssim c' {\sqrt{N}} D_0^{1/\delta}
 \eeq
where $c'$ and $D_0$ are some constants. One can thus estimate the entanglement of the state across any cut to be bounded (\ref{SA})
 \beq
 S_A \alt \frac{k_1}{\delta}  +\frac{1}{2} \log N + k_2
 \eeq
where $k_{1,2}$ are constants.

\section{Constructing states with arbitrarily small energy variance}
\label{sec:construction}

In order to explore
how much entanglement is there in pure states with small energy variance,
we present here two explicit constructions for families of states with well-defined
variance. In the next section we will show that they also have controlled entanglement.

\subsection{Product states}
\label{subsec:product}

We start considering the trivial case of product states
 \beq
 |p\rangle = |p_1\rangle\otimes \ldots |p_n\rangle.
 \eeq
where $p_n$ are single spin states. These states have no entanglement for any bipartition, and their variance reads
 \beq
 \label{deltap}
 \delta_p^2 = \sum_{n=1}^N \sum_{m=-1}^1 \langle p|h_n h_{n+m}|p\rangle
 - \langle p|h_n|p\rangle \langle p|h_{n+m}|p\rangle.
 \eeq
This tells us that $\delta_p\le \sqrt{6N}$. Actually, one can always construct a product state with $\delta \ge y \sqrt{N}$ for some constant $y$,
since the averaged variance over all product states can be easily seen to be $O(N)$.
Thus, $\delta\sim {\sqrt{N}}$ can be obtained with zero entanglement.

We will make use of the following result from~\cite{hartmann2004gauss}.
In the case of a product state $\ket{p}$ with mean energy $E_p=\langle p|H|p\rangle$ and energy variance $\sigma_p^2$ such that $\sigma_p=a\sqrt{N}$, with $a>0$, the local density of states (or energy distribution) converges in the thermodynamic limit to a Gaussian
 \beq
 \label{Hartmann2}
 \rho_p(E) = \frac{1}{\sqrt{2\pi}\sigma_p} e^{-(E-E_p)^2/2\sigma_p^2}.
 \eeq
By local density of states we mean that for any interval $\Delta=[E_1,E_2]\in \sigma(H)$, if we take $\mathds{P}_\Delta$ to be the projector onto the subspace spanned by the eigenstates of $H$ with eigenvalue in that interval, then
\footnote{The result derived in\cite{hartmann2004gauss} states that the local density of states, usually defined as $\rho_p(E)=\sum_n |{\langle {p}} \ket{E_n}  |^2 \delta (E-E_n)$,
converges weakly to the Gaussian form.}
 \beq
 \langle p|\mathds{P}_{\Delta}|p\rangle \simeq \int_{E_1}^{E_2} \rho_p(E) dE.
 \eeq

\subsection{Entangled states}
\label{sec:theory}

We will consider here states of the form
 \beq
 \label{Psip}
 |\Psi\rangle = \frac{1}{\cal N}\sum_{m=-M_0}^{M_0} c_m e^{i2m H/N} |p\rangle
 \eeq
where $p$ is a product state with $E_p=\langle p|H|p\rangle=0$, $\sigma_p=a\sqrt{N}$ with $a=\bigO(1)$, and ${\cal N}$ is the normalization factor.

In the following paragraphs we specify two explicit choices for the
coefficients $c_m$ of the sum in \eqref{Psip}, such that the variance of the resulting state
$\ket{\Psi}$
systematically decreases with the number of terms in the sum $M_0$.

\subsubsection{Cosine filter.}
\label{subsec:cosine}

We consider the following operator
\beq
 \label{series}
 \left[\cos\frac{H}{N}\right]^M =  \frac{1}{2^M} \sum_{m=-M/2}^{M/2}  {M \choose M/2-m} e^{i2mH/N}.
 \eeq
 The sum above can actually be restricted to $-x\sqrt{M}\leq m \leq x\sqrt{M}$,
with the error scaling as a Gaussian function of $x$, so that it can be made arbitrarily small by a judicious choice of $x=\bigO(1)$.
We have indeed checked that taking $x=2$ the relative error is smaller than $10^{-3}$ for $N\le 1000 $ and
$M\leq(100 N)^2$.
 The action of this operator on $\ket{p}$ can thus be written in the form \eqref{Psip} with $M_0=x\sqrt{M}$ terms.

 Using the fact that
  $\cos^M(X)\simeq e^{-MX^2/2}$ for $|X|<1$,
  and the Gaussian form \eqref{Hartmann2} of the local energy density of $\ket{p}$,
the variance of the resulting state is found to be $\delta^2 = (1/\sigma_p^2+2M/N^2)^{-1}$ which
for large enough systems scales as
 \beq
 \delta=\frac{N}{\sqrt{2 M}}.
 \label{eq:deltaCos}
 \eeq

\subsubsection{Chebyshev filter.}
\label{subsec:cheby}

We found that for the numerical implementation, it is more convenient to use the alternative construction
we describe here, which attains a similar scaling, but allows more efficient simulations.

A piecewise continuous function $f(x)$ for $-1\leq x\leq 1$ can be expanded as $f(x)=\sum_{m} p_m T_m(x)$
in terms of Chebyshev polynomials of the first type.
The coefficients of the expansion can be computed using the orthogonality properties of the polynomials as
$p_m=C_m^{-1} \int dx w(x) f(x) T_m(x)$, where $w(x)=(1-x^2)^{-1/2}$ is the weight function for this family of polynomials,
and $C_m$ are the normalization factors,  $C_{m}=\int_{-1}^{1}  dx w(x) T_n(x)^2$, namely $C_{m>0}=\pi/2$
and $C_{0}=\pi$.
The truncation of the sum to a finite $M$ provides an approximation to the function, which
 exhibits characteristic (Gibbs) oscillations near a discontinuity.
The kernel polynomial method\cite{kpm2006} reduces this effect and improves the convergence of the truncated series
by multiplying the coefficients by specific factors $g_m^{(M)}$.
In particular, we use the Jackson kernel, for which
\beq
g_k^{(M)}=\frac{(M-k+1) \cos\frac{\pi k}{M+1}+\sin \frac{\pi k}{M+1}\cot \frac{\pi}{M+1}}{M+1}.
\eeq

In the case of the Dirac delta function, all coefficients for odd polynomials vanish, and the $M-$th order approximation using the KPM reads
\beq
\delta(x)\approx \sum_{n=0}^{\lfloor M/2\rfloor}
(-1)^n \frac{2-\delta_{n0}}{\pi}
 g_{2n}^{(M)}T_{2n}(x).
\label{eq:chebypol}
\eeq

By applying the same truncated series to the rescaled Hamiltonian $H/N$,
we obtain the operator
\beq
O_M=\sum_{m=0}^{\lfloor M/2 \rfloor} (-1)^m \frac{2-\delta_{m0}}{\pi} g_{2m}^{(M)} T_{2m}\left(\frac{H}{N}\right ).
\label{eq:chebydelta}
\eeq

The result  of applying this operator to the product state $\ket{p}$ can also be written in the form
\eqref{Psip}.
The Chebyshev polynomials fulfil $T_n(\cos x)=\cos(nx)$. Then
\begin{align}
T_n \left( \frac{\alpha {H}}{N}\right)&\approx T_n \left[\sin \left(\frac{\alpha {H}}{N}\right)\right]
=T_n \left[\cos \left(\frac{\pi}{2}-\frac{\alpha {H}}{N}\right)\right]\nn \\
&=\cos\left(n\frac{\pi}{2}-n\frac{\alpha {H}}{N}\right)\nn \\
&=\frac{1}{2}\left[(-i)^n e^{i\alpha n {H/N}}+i^n e^{-i\alpha n {H/N}}\right].
\end{align}
Choosing a constant $\alpha \ll 1$ and rescaling $H \to \alpha H$ ensures that the approximation
holds for all eigenvalues of the argument.
In the case we study here it is enough to consider $\alpha=1$, since the contributions of large
eigenvalues (for which in principle the relation may fail)
are rapidly suppressed.

So, finally we can write
\beq
O_M\ket{p}\sim \sum_{m=-\lfloor M/2 \rfloor}^{\lfloor M/2 \rfloor} K_m e^{i 2m H/N} \ket{p}.
\label{eq:chebysumexp}
\eeq
which is of the form \eqref{Psip} with $M_0=\lfloor M/2 \rfloor$ terms in the sum.

The truncated sum \eqref{eq:chebypol} actually approximates a Gaussian\cite{kpm2006}  $e^{-x^2/(2\sigma^2)}$
with variance $\sigma \sim \pi/M $.
Similarly, the corresponding operator series \eqref{eq:chebydelta} approximates
$O_M \sim e^{-(H M/\sqrt{2}\pi N)^2  }$.
This fact, combined with \eqref{Hartmann2}, enables us to evaluate the variance of the state
 $O_M \ket{p}$,
 \beq
\delta^2=(1/\sigma_p^2+ 2 M^2/\pi N^2)^{-1} \sim \left (\frac{\pi N}{\sqrt{2} M}\right)^2
\label{eq:deltaCheby}
\eeq
where the last step results from considering the limit $N\gg1$.

\section{Relation between entanglement and energy variance}
\label{sec:bounds}

The entanglement of any state with the form \eqref{Psip} can be upper bounded by a function of $M_0$ and $N$.
Using a result from \cite{hartmann2004gauss}, in the thermodynamic limit,  the overlap between two terms in the sum, $m$ and $m'$
decreases with their separation  as
 \beq
 \label{Hartmann}
 \left|\langle p | e^{i 2 (m-m') H/N} |p\rangle \right|^2 \simeq e^{-[2 (m-m') \sigma_p/N]^2}.
 \eeq
Terms for which the separation $m-m' \ll N/(2\sigma_p)=\sqrt{N}/(2 a)$ are almost
proportional to each other.
Therefore we can reduce the number of terms in the sum by a factor $\sqrt{N}$ by defining a constant $\gamma \gg a $ and grouping each set of  $\sqrt{N}/\gamma$ consecutive terms, as
\beq
\ket{\Psi}\simeq \frac{1}{\mathcal{N}} \sum_{k=-\gamma M_0/\sqrt{N}}^{\gamma M_0/\sqrt{N}} C_k e^{2 i k  H/{\gamma \sqrt{N}}} \ket{p},
\eeq
where $C_k=\sum_{m=(k-1)\sqrt{N}/\gamma}^{k\sqrt{N}/\gamma-1} c_m $.

The entanglement generating capacity of a local Hamiltonian as  \eqref{H} is bounded\cite{duer2001ent,Bennet2003,Bravyi2007,acoleyen2013rate},
so that, when acting with the evolution operator $e^{ir H}$ on a product state,
the entanglement for any given cut can only increase linearly with $r$, independently of $N$.

Thus, each (normalized) term $e^{2 i k H/{\gamma \sqrt{N}}} \ket{p}$ appearing in the sum can be approximated by a MPS
with bond dimension $D_1^{2 |k|/{\gamma \sqrt{N}}}$, for some $D_1=\bigO(1)$.
Since the bond dimension of a sum of MPS  is at most the sum of the bond dimensions of its terms,
 $\ket{\Psi}$ can be approximated by a MPS with
\beq
D\leq\sum_{k=-\gamma M_0/\sqrt{N}}^{\gamma M_0/\sqrt{N}} (D_1^{2 /{\gamma \sqrt{N}}})^{|k|}
\label{eq:Dmax}
\eeq
or, in the limit of large system size $N\gg1$,
\beq
D\leq  \frac{\gamma \sqrt{N}}{ \log D_1}\left ( D_1^{2 M_0/N}-1 \right),
\label{eq:boundD0}
\eeq

From equations  \eqref{eq:deltaCos} and \eqref{eq:deltaCheby},
we see that the energy variance in both constructions presented in the previous sections
decreases precisely as
$\delta\propto N/M_0$, so that in \eqref{eq:boundD0}
we may substitute $D_1^{2 M_0/N}=D_0^{1/\delta}$,
with a value $D_0$, specific for each case, that absorbs the corresponding constant exponents.
Then the bond dimension in both cases scales (for large systems) as
\beq
D\leq   \frac{\gamma \sqrt{N}}{ \log D_1}\left ( D_0^{1/\delta}-1 \right),
\label{eq:boundD}
\eeq
which is one of our main results.
The expression reduces to \eqref{bonddim} when $\delta\ll 1$.

Correspondingly, using \eqref{SA}, in the large $N$ limit, the entanglement is bounded by
\beq
S\leq  \log \left( D_0^{1/\delta}-1\right)+\frac{1}{2}\log N+k_2,
\label{eq:boundS}
\eeq
which reduces to $S\leq k_1/\delta +\log \sqrt{N} +k_2$ for small enough $\delta$.

\subsection{Implications and limitations}
\label{subsec:discussion}

Let us briefly comment the conditions and consequences of the results derived above.

Regarding the conditions in our derivations:
\begin{enumerate}[label=(\roman*)]
\item \label{cond.1}
Equations \eqref{eq:boundD} and \eqref{bonddim} have to be taken as upper bounds: the estimations we made to compute the bond dimension of the approximations to the state (\ref{Psip}) consider a worst-case scenario, where the Hamiltonian $H$ generates as much entanglement as possible, and the bond dimension of a linear combination of states is the sum of the bond dimensions of each of them.
\item \label{cond.2}
The constructions introduced in this section prepare states of a given variance, and we have shown that they possess bounded entanglement,
but there may well be other states with less entanglement for the same variance.
Indeed, we cannot expect to find tight bounds, since it is possible to construct examples of exact product states in the middle of the spectrum for specific interacting systems. For instance, just take a staggered ferromagnetic Hamiltonian, $H=\sum_n (-1)^n \vec\sigma_n \vec \sigma_{n+1}$, for which any product state $|p\rangle^{\otimes N}$ is an eigenstate with zero energy (for $N$ even).

In order to try to obtain better general bounds,
one could separate $H=H_L+H_R+h$, where $H_L$ ($H_R$) is the part of $H$ acting on the left (right) half of the chain, and $h$ the one that connects them.
Then, the state $|\Psi\rangle = |\varphi_L\rangle\otimes|\varphi_R\rangle$, where $\ket{\varphi_{L,R}}$ are the ground eigenstates of $H_L$ and $-H_R$, respectively, would
have a variance of $O(1)$, as only $h$ contributes to it.
Furthermore, if $H_L$ and $-H_R$ are gapped, then the states $\ket{\varphi_{L,R}}$  satisfy an area law, and thus the entropy of $|\Psi\rangle $ along any cut is
upper-bounded by a constant independent of $N$, which improves the scaling in \eqref{eq:boundS}.
However, if $H_L$ or $-H_R$ are gapless, the bound on the entropy scales again with $\log N$, possibly with a larger prefactor than  in \eqref{eq:boundS}.
Even in the gapped case, the bond dimension of $|\Psi\rangle$ is not bounded,
but the states $\ket{\varphi_{L,R}}$ can be approximated with MPS $\ket{\phi_{L,R}}$ of bond dimension
$D=\exp[O(\log(N)^{3/4} \epsilon_{L,R}^{-1/4})]$ \cite{arad2013area,arad2017},
where $\epsilon_{L,R}=\|\ket{\varphi_{L,R}}-\ket{\phi_{L,R}}\|^2$.
In that case, the bond dimension of the state $|\Phi\rangle=|\phi_L\rangle\otimes |\phi_R\rangle$ would be upper bounded by a sublinear function of $N$,
but its variance may be as large as $\epsilon N^2$: just taking $|\phi_L\rangle = \sqrt{1-\epsilon_L} |\varphi_L\rangle + \sqrt{\epsilon_L} |\phi'\rangle$, with $\phi'$ the maximally excited state of $H_L$,
even letting $\epsilon=1/\log N$ (which would cost $D=O[{\rm poly}(N)]$, as we obtained above), the variance would only  be upper bounded by $\bigO(N^2/\log N)$.

\item \label{cond.3}
We have shown that, in the limit $N\gg1$, our constructions yield a variance of the form $\delta\sim N/M_0$, for $M_0$ terms in the sum, but for finite systems
we expect some corrections to appear.
In particular, to make the derivation of the cosine operator rigorous, we should scale $x$  and $\gamma$ with $N$ and $\delta$; however, since the error we made by truncating the series (\ref{series}) is exponentially small in $x$, the corrections will only depend logarithmically on those quantities.

\item \label{cond.4}
Although we have described how to use our construction to obtain states in the middle of the spectrum, with
$E\simeq 0$, it can also be used for other energies $E_0$,
as long as there exists a product state with mean energy $\bra{p}H\ket{p}=E_0$ on which we can apply the filter, after replacing $H\ra H-E_0$.
In particular for qubits, as considered here, such an initial product state can be shown to
exist for any energy  $|E_0|\le N h_{\rm min}/6$.
This can be seen as follows.
First, for each odd term $h_{2n-1}$ in $H$, we define
\beq
m_{2n-1}:=\max_{\ket{\varphi\,\phi}} \left | \bra{\varphi\,\phi} h_{2n-1} \ket{\varphi\,\phi}\right |,
\eeq
where the maximization is over all product states, and we define local unitaries $U_n$ such that
$m_{2n-1}=  \bra{00}\tilde{h}_{2n-1}\ket{00} $, where
$\tilde{h}_{2n-1}$ is ${h}_{2n-1}$ conjugated with $U_{2n-1}\otimes U_{2n}$.
It is easy to show that $m_{2n-1}\geq h_{\min}/3$, where the 
 bound is tight (for instance, for $h_{2n-1}=\vec\sigma_{2n-1} \vec\sigma_{2n}$).
If we conjugate $H$ with the product of all $U_n$, we obtain
 \beq
 \label{eq:maxim}
 \tilde H = \sum_n \left ( a_n \sigma^z_n \sigma^z_{n+1} + b_n \sigma^z_{n} \right )+ H',
 \eeq
where $|a_{2n-1}|+|b_{2n-1}|=m_{2n-1}$, and $H'$ does not contain terms of the form: $\sigma^z_{2n-1},\sigma^{z}_{2n-1}  \sigma^z_{2n}$, $\sigma^{z}_{2n-1}  \sigma^{x,y}_{2n}$.
 The first two cannot appear since they are already included in (\ref{eq:maxim}) and ${\rm tr}_n h_n=0$, so they cannot come from $h_{2n}$. The terms $\sigma^{z}_{2n-1}  \sigma^{x,y}_{2n}$ cannot appear either, as if they do, then there would exist a product state such that the expectation value of $\tilde{h}_{2n-1}$ would be larger than the maximum, $m_{2n-1}$. 
Now, we can take a product state of the form $\otimes_i\ket{p_i}$, where we choose $p_i\in\{0,\,1\}$ from left to right making sure that all expectation values give a positive value, 
and we call, for the even terms, $m_{2n}:= \bra{p_{2n} p_{2n+1}} \tilde h_{2n} \ket{p_{2n} p_{2n+1}}\geq0$. 
Then the corresponding energy is $\sum_{\ell} m_{\ell} \ge \sum_n m_{2n-1} \ge N h_{\rm min}/6=E_0$. Similarly, we can construct product states with energy $-E_0$.
\end{enumerate}

Regarding the consequences:
\begin{enumerate}[label=(\roman*)]
\item \label{conseq.2}
Eq.  \eqref{eq:boundD} implies that there are states with constant energy variance and with a bond dimension that only scales polynomially with $N$. Those states have at most $\log N$ entanglement along any cut, and thus only slightly violate the area law.
Conversely, if we keep the bond dimension constant, the variance must increase with $\delta\sim\sqrt{N}$.

 \item \label{conseq.3}
It is possible to build states with a variance decreasing as $\delta\sim 1/\log N$ but still keeping a polynomial bond dimension.
Notice however that this case may be affected by the corrections mentioned above. For instance, in the cosine construction in \ref{subsec:cosine}, if
the factor $x$ needs to grow logarithmically with $N$,  $\delta$ would have to decrease as some power of $1/\log N$.

 \item \label{conseq.4}
For the state \eqref{Psip} to reproduce the thermal properties in the thermodynamic limit (i.e. for the expectation values of all local observables -- with any support -- to converge to their thermal values)
 the state needs to have an extensive value of the entanglement entropy, and thus $\delta \le 1/N$.
However, if one is only interested in the observables in a region of fixed size $L$, it may be enough that the entropy of that region is $L$, so that keeping $\delta$ constant (or slightly decreasing with $N$), may be enough to locally thermalize for sufficiently large $N$, and in this case the local temperature may vary on length scales much larger than $L$\cite{dymarsky2019canuni}.
Notice that we can also apply the arguments leading to \eqref{eq:boundS} to a subsystem  in the middle of the chain, by simply considering a double chain obtained by \emph{folding} the original one in two around the center of the subsystem. Thus a similar bound  \eqref{eq:boundS} (with different constants) holds for the entropy of the subchain of length $L$.
\end{enumerate}

In the next sections we investigate numerically all the points raised above. We give an explicit construction to build the MPS, and numerically check to what extent  \eqref{eq:deltaCheby}, \eqref{eq:boundD} and \eqref{eq:boundS} are obtained for moderate values of $N$ and $\delta$.
We also explore how close to thermal are the reduced density matrices of these states for small subsystems.


\section{Numerical results}
\label{sec:num}

\subsection{Numerical implementation}
\label{subsec:algo}

In order to achieve a fixed variance $\delta^2$, both
constructions presented in the previous section would require a number of
states in the sum \eqref{Psip} proportional to $N/\delta$.
To implement the first strategy \ref{subsec:cheby} numerically we can use
standard MPS time evolution techniques in order to approximate each term of the sum
starting from the product state $\ket{p}$.
We have observed, nevertheless, that the truncation error accumulates fast
when compressing the terms of the sum.
Iteratively applying the cosine operator $\exp[i H/N]+\exp[i H/N]$
produces numerically more stable results, but requires
$(N/\delta)^2$ iterations.
A more efficient strategy is to implement the Chebyshev construction from \ref{subsec:cheby}
by approximating with a MPS the action of each term in (\ref{eq:chebydelta}) on the initial state.

Starting from the product state, $\ket{p}$, we construct the terms of the sum (\ref{eq:chebydelta}) as follows.
The first two Chebyshev polynomials are exact matrix product operators\cite{pirvu10mpo} (MPO) with small bond dimension,
$T_0(\tilde{H})=\Id$, and $T_1(\tilde{H})=\tilde{H}$, where $\tilde{H}=\alpha H/N$ is the properly rescaled Hamiltonian,
so that its spectrum is within the convergence domain of the Chebyshev expansion.
In practice, to choose $\alpha$ it is enough to obtain an estimate of the edges of the spectrum, $E_{\min}$ and $E_{\max}$, which can
be done efficiently using DMRG, and then to take $\alpha<N/\max(|E_{\min}|,|E_{\max}|)$, where the inequality is guaranteed by fixing for instance
$\alpha$ as $0.9$ times the rhs).

The 0-th order approximation corresponds to $\ket{\Psi_0}=g_0^{(M)} c_0 \ket{p}$.
Then, we iterate until the desired order $M$,
\beq
\ket{\Psi_{n+1}}=\ket{\Psi_{n-1}}+g_n^{(M)} c_n T_n(\tilde{H}) \ket{p},
\eeq
where $c_n$ can be read from \eqref{eq:chebydelta}.
In principle, the higher order polynomials could be also computed as MPO using the recurrence
$T_{n+1}(\tilde{H})=2\tilde{H}T_n(\tilde{H})-T_{n-1}(\tilde{H})$
using the standard algorithms.
However, since  only the action of each polynomial on the initial state, $\ket{T_n(p)}:=T_n(\tilde{H}) \ket{p}$, appears in the sum,
it is  more efficient to compute these vectors, which satisfy the same recurrence relation
\beq
\ket{T_{n+1}(p)}=2 \tilde{H} \ket{T_n(p)}-\ket{T_{n-1}(p)},
\eeq
and use them to update the sum  as $\ket{\Psi_{n+1}}=\ket{\Psi_{n-1}}+g_n^{(M)} c_n \ket{T_{n}(p)}$.
This allows us to operate directly with MPS, avoiding the more costly operators.
Notice that the use of Chebyshev polynomials of the Hamiltonian combined with tensor network techniques
was first suggested in\cite{holzner2011chebyMPS} for approximating spectral functions
and has been later used for time evolution\cite{wolf2015cheb,Halimeh2015cheby,xie2018cheb} and density of states calculations~\cite{Yang2019chebydos}.
The technical details involved in our computation of the intermediate $\ket{T_{n+1}(p)}$ vectors are virtually the same described in those references.

\subsubsection{Models and initial states.}
We consider two spin $1/2$ quantum chains to probe our construction, namely the Ising model with longitudinal and transverse field, and the XYZ
Hamiltonian in a magnetic field,
\begin{align}
\His&=J \sum_i \sigma_z^{[i]}\sigma_z^{[i+1]}+g  \sum_i \sigma_x^{[i]}+h  \sum_i \sigma_z^{[i]},
\label{eq:Hising}
\\
\Hxyz&= \sum_i \left (J_x\sigma_x^{[i]}\sigma_x^{[i+1]}+J_y\sigma_y^{[i]}\sigma_y^{[i+1]}+J_z\sigma_z^{[i]}\sigma_z^{[i+1]}\right ) \nn \\
& + h  \sum_i \sigma_z^{[i]}.
\label{eq:Hxyz}
\end{align}
We consider non-integrable points, and choose the sets of parameters ($J=1$, $g=-1.05$, $h=0.5$) for $\His$
and ($J_x=1.1$, $J_y=-1$, $J_z=0.9$, $h=1.2$) for $\Hxyz$.

In both cases, we select a product initial state that has energy close to $0$. In the case of the Ising model, we use
$\ket{p}=\ket{Y+} := \left(\frac{\ket{0}+ e^{i\pi/2} \ket{1}}{\sqrt{2}}\right)^{\otimes N}$, which has $\bra{Y+} \His \ket{Y+}=0$
for every system size.
In the case of the XYZ chain, we use a staggered configuration $\ket{p}=\ket{\Z2} := \left(\ket{0}\ket{0} \ket{1}\ket{1}\right)^{\otimes N/4}$
(if the chain length is even, but not a multiple of 4, the last pair of sites is in $\ket{0}$),
with constant energy $J_z$ (for even chain lengths), and thus vanishing energy density in the thermodynamic limit.  
Note that these initial states are spatially homogeneous on large length scales, so that as they approach local thermal equilibrium at large $M$, the local temperature is spatially uniform.  Below, in section~\ref{sec:inhom}, 
we also explore spatially nonuniform initial states.

Using the method described above, we
 compute the Chebyshev sum to $M$-th order for different values of $M$,
and for system sizes between $N=20$ and $100$. In each case, we allow bond dimensions between $D=200$ and $1000$
(notice that $D=1024$ is exact for our smallest system $N=20$).

\subsection{Scaling of the variance}
\label{subsec:numvar}

\begin{figure}[h]
\centering
\subfloat[Ising model: variance vs. $M$]{\label{fig:cheby_Delta2vsN_Ising}\includegraphics[width=.8\columnwidth]{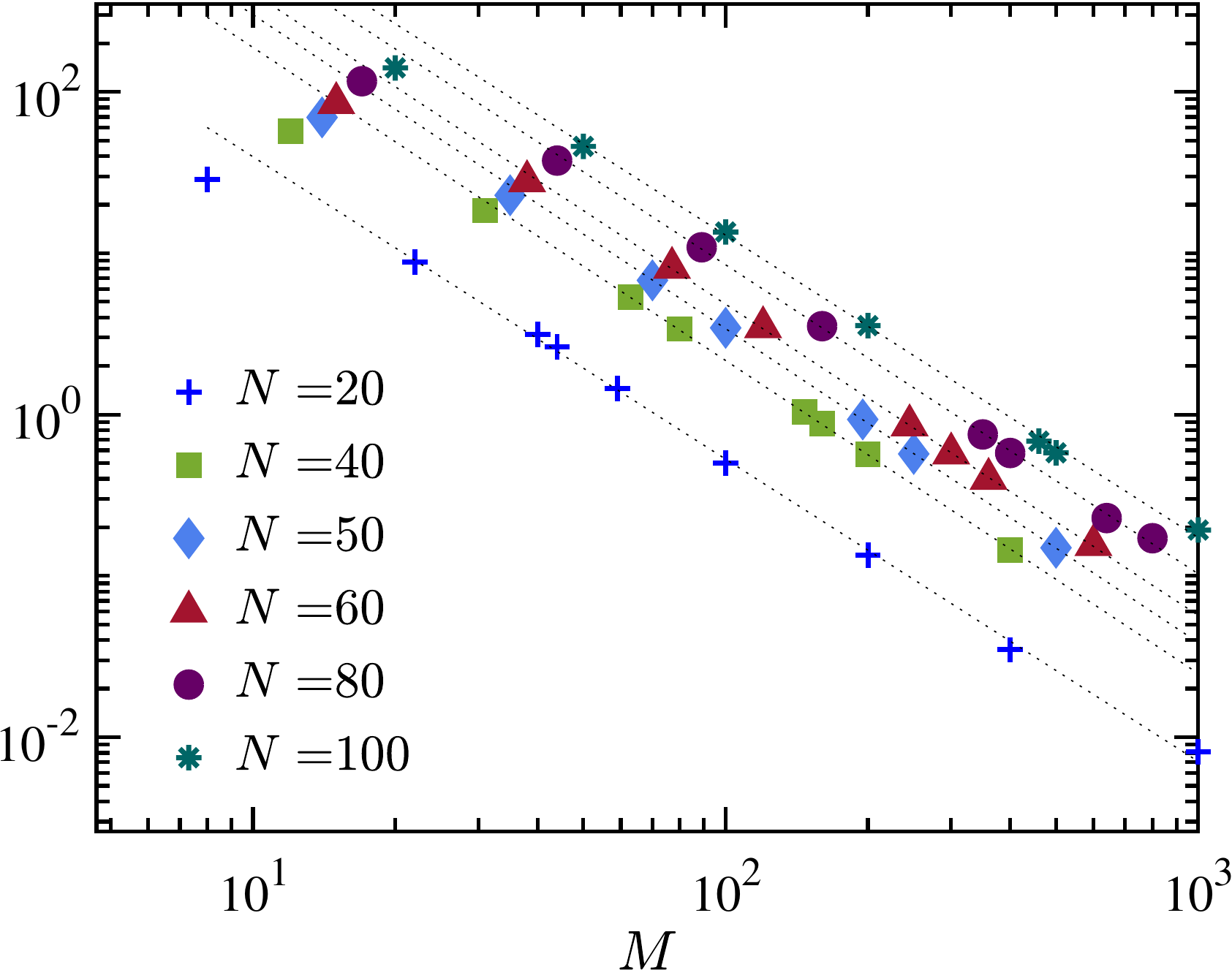}}\\
\subfloat[XYZ model: variance vs. $M$]{\label{fig:cheby_Delta2vsN_Heis}\includegraphics[width=.8\columnwidth]{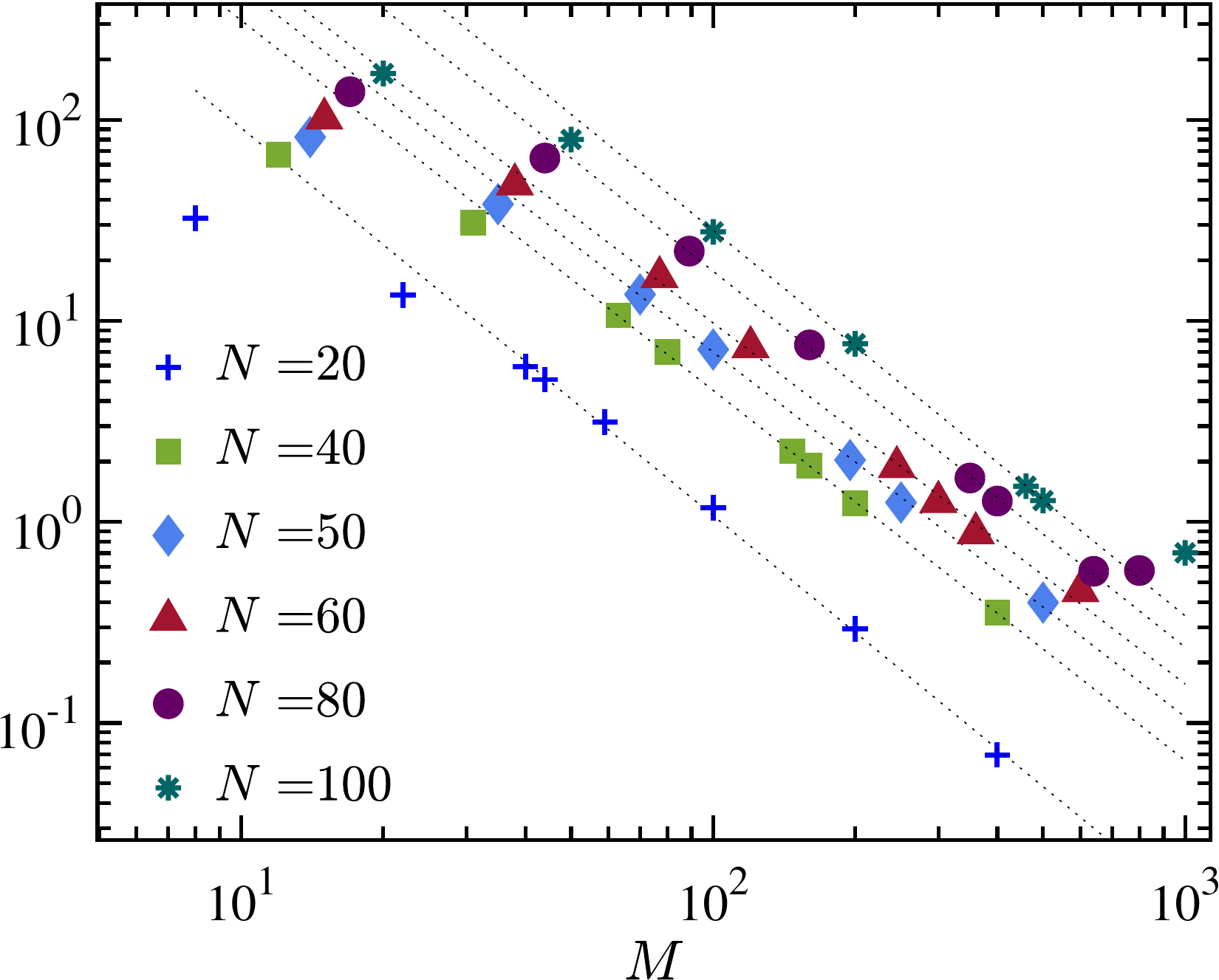}}\\
\caption{Scaling of the (final) variance $\delta^2$ as a function of the order $M$ of the Chebyshev expansion,
for different system sizes $N=20-100$ in the Ising (upper) and XYZ (lower panel) setups.
We show the results for the largest bond dimension used $D=1000$. For the largest $M$,
the results are not yet converged, and they deviate from the straight line (specially noticeable for the XYZ case).
Before truncation limits the decrease of the variance, the scaling is consistent with the expected $\delta^2\propto 1/M^2$ (up to small $M$ corrections).
The dotted lines show a power law fit for the intermediate (converged) points of each size, at the largest bond dimension,
with exponents for $\delta^2$ between -1.95 and -1.88 (Ising) and between -1.93 and -1.80 (XYZ).
}
\label{fig:cheby_Delta2vsM}
\end{figure}

\begin{figure}[h]
\centering
\subfloat[Ising model]{\label{fig:cheby_Delta2vsN_Ising}\includegraphics[width=.48\columnwidth]{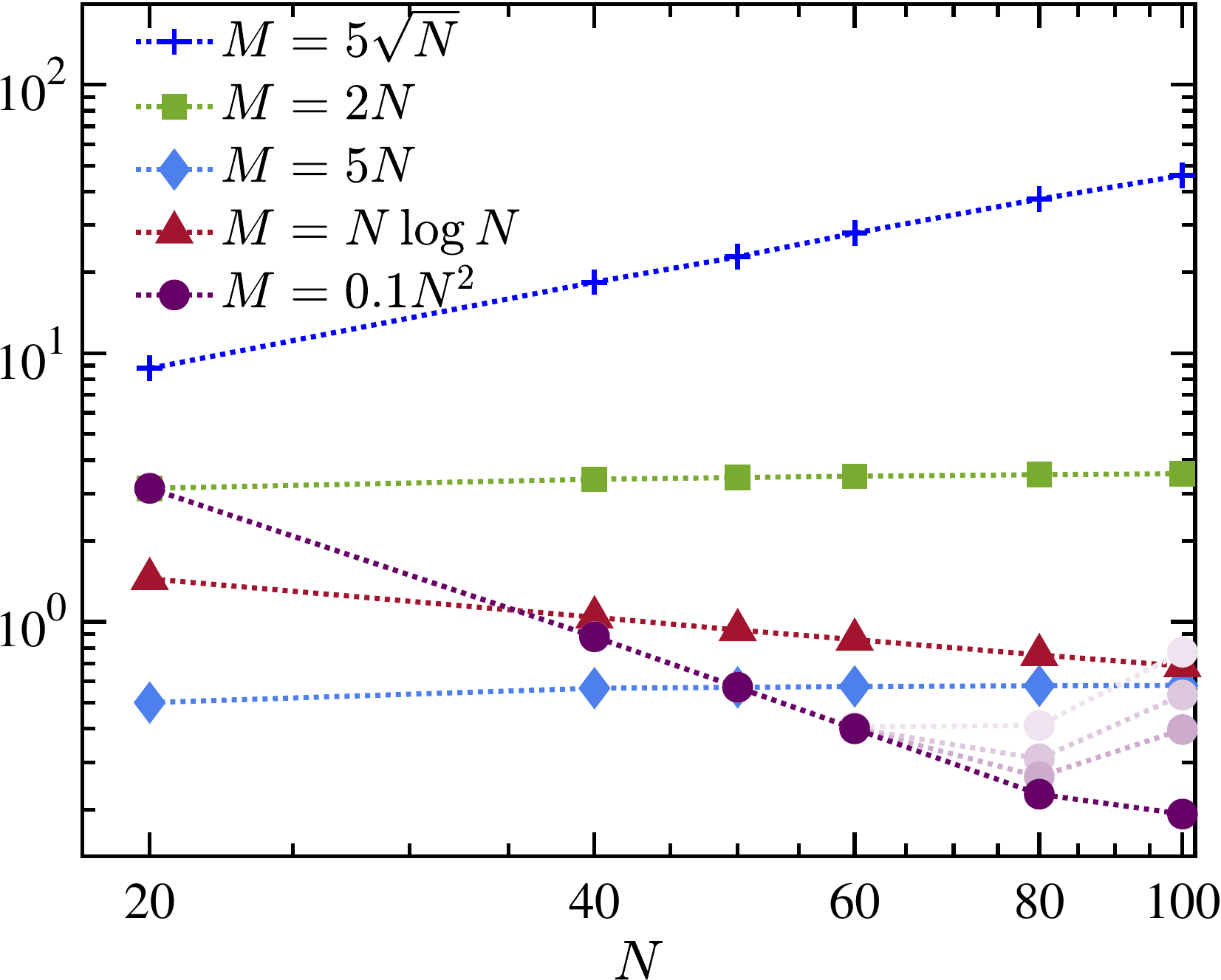}}
\hspace{0.1pt}
\subfloat[XYZ model]{\label{fig:cheby_Delta2vsN_Heis}\includegraphics[width=.48\columnwidth]{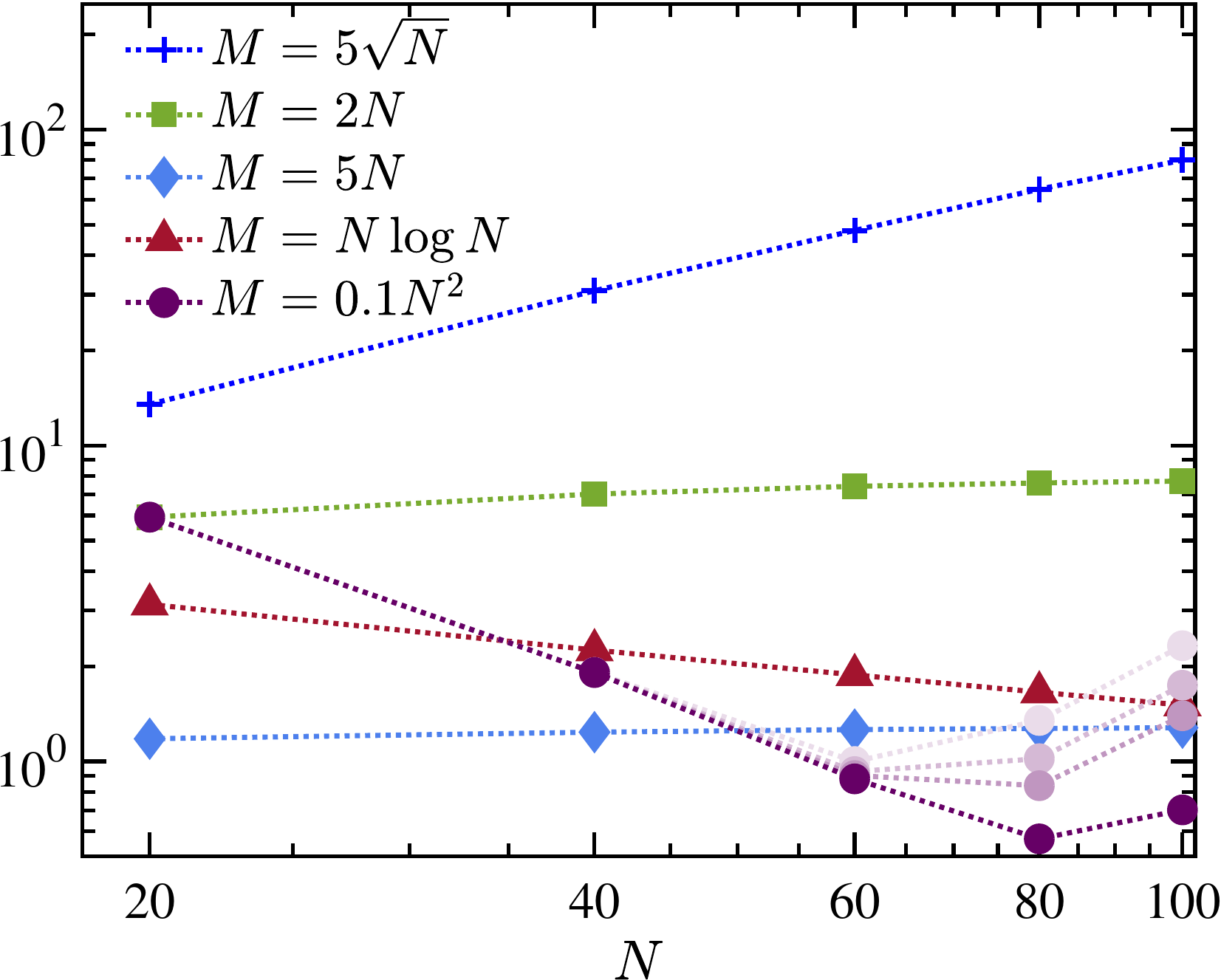}}
\caption{ Final variance $\delta^2$ as a function of the system size $N$, for number of steps corresponding to various functions $M=f(N)$,
in the Ising (left) and XYZ (right pannel) setups.
We show the results with the largest $D=1000$, converged for all sets except $M\propto N^2$ (purple circles), for
which we show in lighter shades the results for $D=300-500$ (lighter to darker).
}
\label{fig:cheby_Delta2vsN}
\end{figure}

To probe whether the energy variance decreases with the number of
terms in the sum according to the asymptotic behavior in \eqref{eq:deltaCheby} already for
the moderate system sizes accessible to the numerics,
we run the procedure described above for both models
using truncation parameters $M$ that vary with the system size.
As shown in figure~\ref{fig:cheby_Delta2vsM},
the scaling is close to the asymptotic one, as  long as the truncation error is not important.
Fitting the final variance for the converged calculations for each system size
to $\delta = A M^{-\eta}$, we find $\eta$ close to $1$ for all cases, with only small deviations
 (see caption of figure~\ref{fig:cheby_Delta2vsM}).

Another way to check the behavior predicted in~\ref{sec:theory} is to examine the scaling of the final variance when the truncation order scales as a certain
function of the system size $M=f(N)$ (see figure~\ref{fig:cheby_Delta2vsN}).
Since we expect $\delta\propto N/M$, a linearly growing $M\propto N$ should maintain constant variance for increasing system size.
Our simulations show that this is practically the case,
and the behavior is consistent in both models.
Nevertheless, we observe a slight increase $\delta \lesssim N^{0.03}$ for Ising (fig.~\ref{fig:cheby_Delta2vsN_Ising})
and $\delta \lesssim N^{0.05}$ for XYZ (fig.~\ref{fig:cheby_Delta2vsN_Heis}).
If $M$ grows faster than linear (see the figure for $M\propto N \log N$ and $M\propto N^2$), the variance decreases with increasing system size.
In the first case, the results are compatible with a descent $\delta \sim 4/\log N$, but also with a power law $N^{-0.223}$ (Ising),
respectively $\delta \sim 1.53/\log N$ or $N^{-0.22}$ (XYZ).
In the quadratic case, the variance drops much faster, compatible with $\delta \propto N^{-0.97}$ (Ising) and $N^{-0.88}$ (XYZ)
within the converged range of sizes (notice that in that case the largest
system size is not converged even with $D=1000$).

\subsection{Scaling of entropy and truncation error}
\label{subsec:entropy}

\begin{figure}[h]
\centering
\subfloat[Ising model]{\label{fig:cheby_S_funN_Ising}\includegraphics[width=.48\columnwidth]{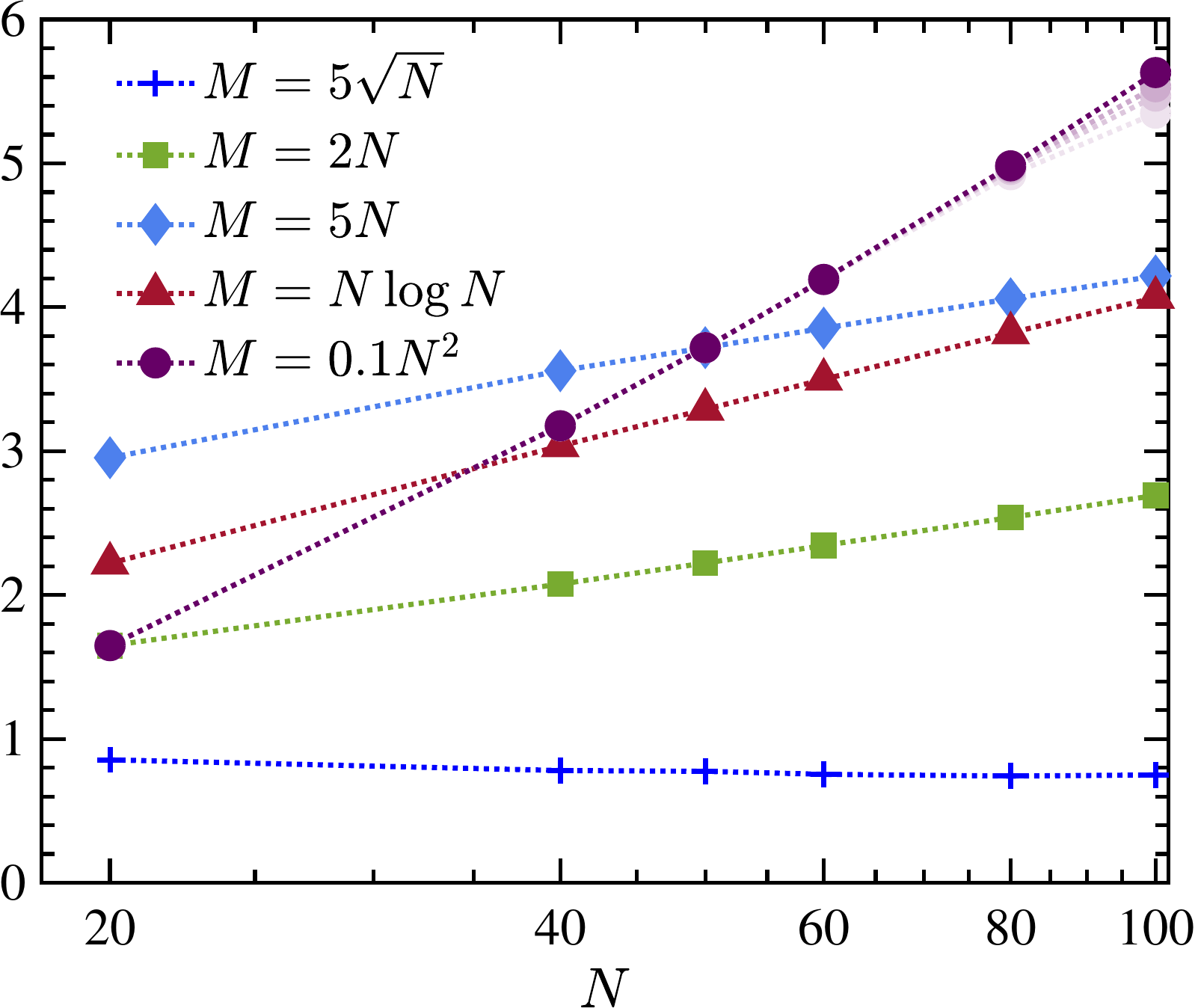}}
\hspace{0.1pt}
\subfloat[XYZ model]{\label{fig:cheby_S_funN_Heis}\includegraphics[width=.48\columnwidth]{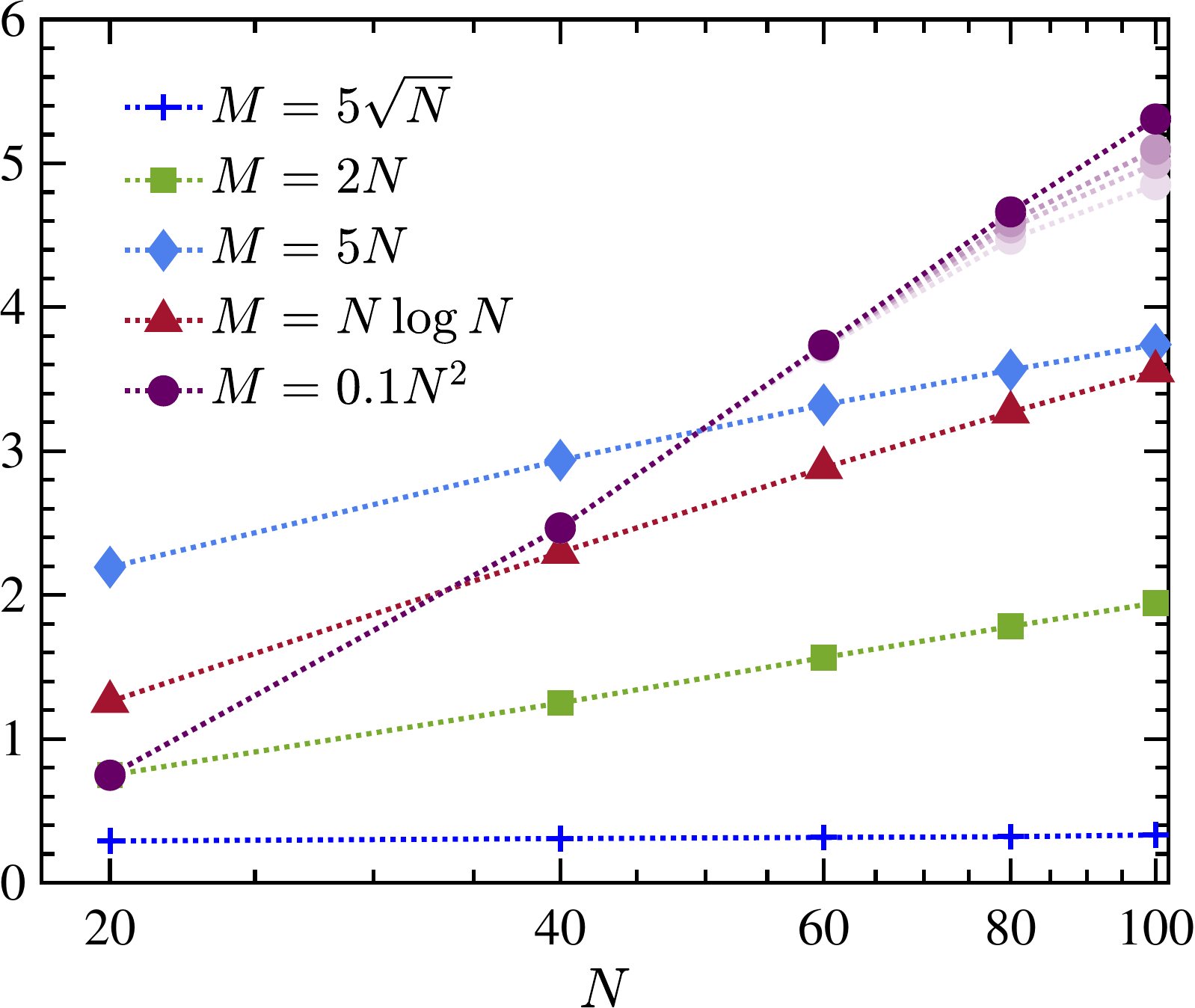}}
\caption{Entropy of the half chain as a function of the system size $N$, after applying the Chebyshev filter 
for different truncations $M=f(N)$ (indicated by the symbols) in the Ising (upper) and XYZ (lower pannel) setups,
for bond dimension $D=1000$.
In the only non-converged case, the lighter symbols show also results for $D=300-500$ (lighter to darker shades).
We observe that almost all cases grow with $\log N$,
except for the case $M\propto \sqrt{N}$, in which the entropy seems upper bounded by a constant.
In the case $M=0.1 N^2$, in which the values are also compatible with an additional linear term in $N$.
}
\label{fig:cheby_S_funN}
\end{figure}

\begin{figure}[h]
\centering
\subfloat[Ising model]{\label{fig:cheby_SvsDelta_Ising}\includegraphics[width=.85\columnwidth]{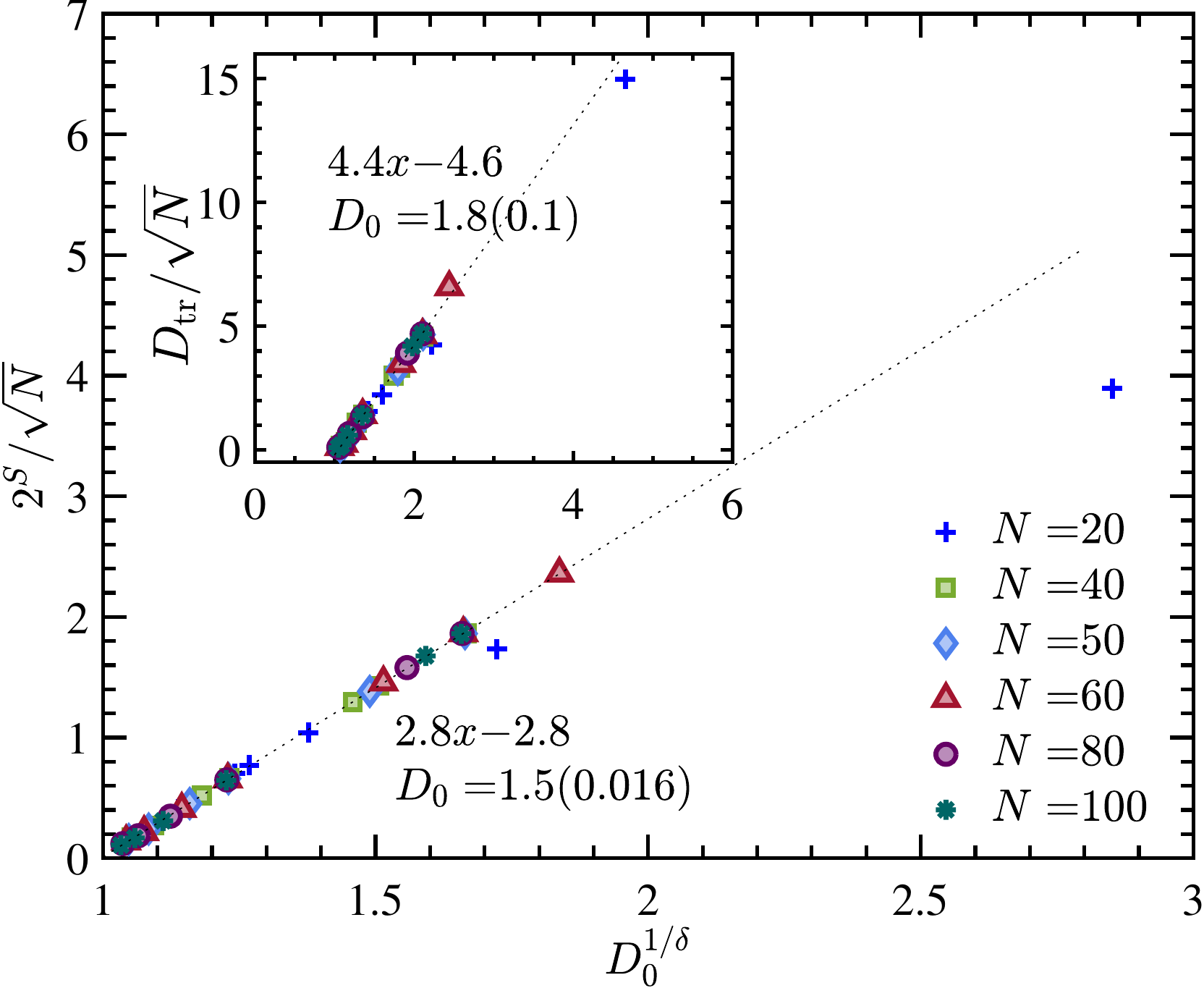}}\\ 
\subfloat[XYZ model]{\label{fig:cheby_SvsDelta_Heis}\includegraphics[width=.85\columnwidth]{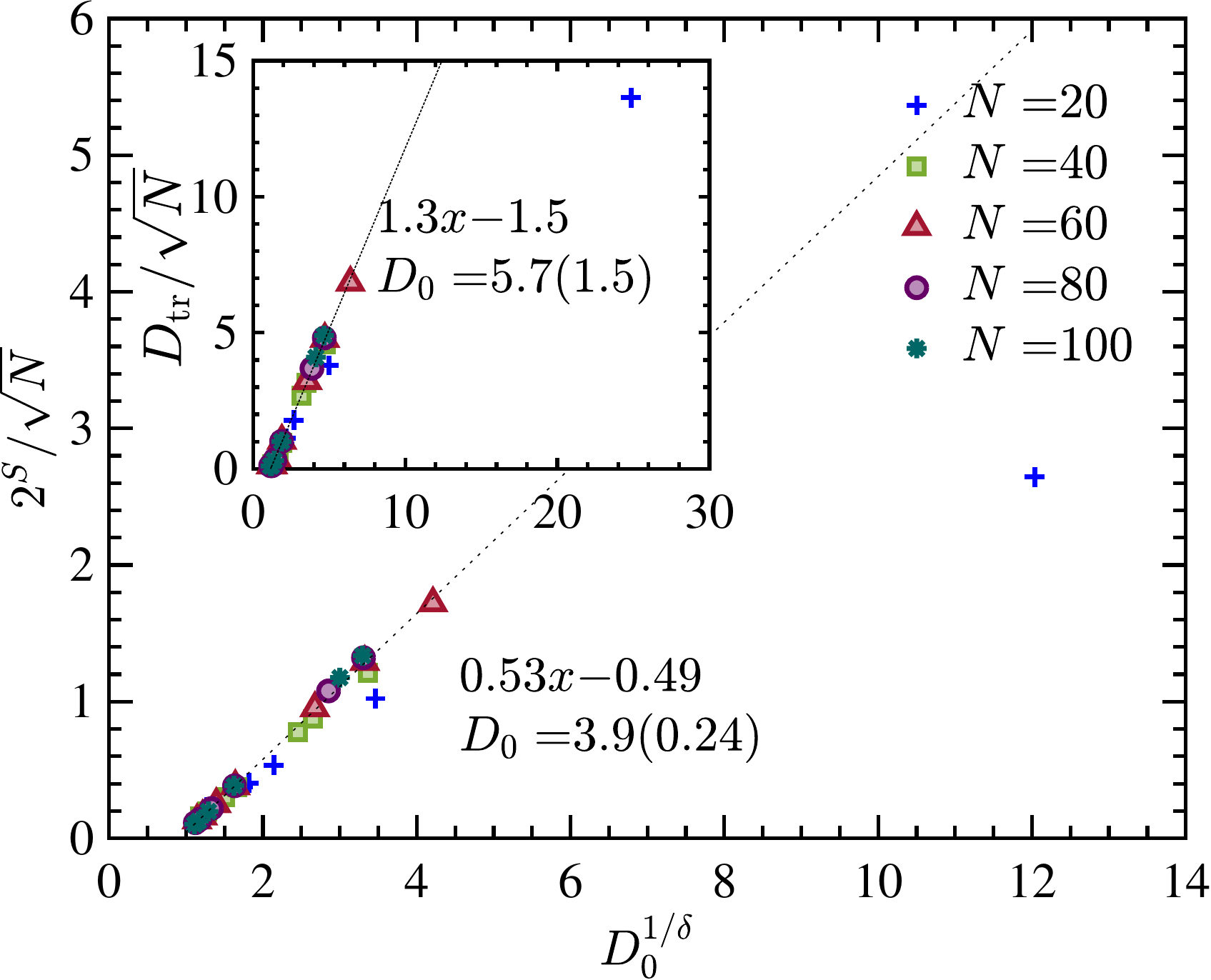}} 
\caption{
Scaling of the exponential of the entropy as predicted by \eqref{eq:boundD}. Except for the smallest systems $N=20$, the
exponential of the half-chain entropy is approximately proportional to $ \sqrt{N} \left( D_0^{1/\delta}-1 \right )$ with a constant $D_0$ (shown in each panel), found from
the individual fits for different system sizes. The dotted line represents the linear fit $2^S/\sqrt{N}= A D_0^{1/\delta}+B$,
and the parameters  of each fit are shown in the corresponding panel.
We observe that the XYZ case (below) exhibits larger deviations for finite systems.
}
\label{fig:cheby_SvsDelta}
\end{figure}

Equations  \eqref{eq:boundD} and \eqref{eq:boundS}  estimate the upper bounds of
the scaling of the entanglement entropy and the bond dimension
with system size for each set of states defined by a function $M=f(N)$. In particular, if  $1/\delta$
scales as $\log N$ or slower, which is the case for all functions discussed above except $M\propto N^2$,
 the bound scales asymptotically as $\log N$.

We check this prediction plotting the numerical results vs. the logarithm of the system size, as shown in
figure \ref{fig:cheby_S_funN}.
Our data show that in almost all cases, the growth of the entropy is compatible with $\log N$.
We fit the resulting entropies  for the largest bond dimension (discarding the smallest system sizes)
for each model (fig.~ \ref{fig:cheby_S_funN_Ising}, \ref{fig:cheby_S_funN_Heis}),
 and find
the following forms~\footnote{In the XYZ case, size $N=50$ is also excluded from the fits, as, different from the others, it is not multiple of 4 and exhibits qualitatively different behavior.}.

\begin{center}
\begin{tabular}{l |c|c }
& Ising  & XYZ\\
\hline
$S_{5\sqrt{N}}$ & $-0.04 \log_2 N+1$ & $-0.003 \log_2 N+0.34$ \\
$S_{2 N}$ & $ 0.46 \log_2 N-0.4 $ & $ 0.47 \log_2 N-1.15 $  \\
$S_{5 N}$ & $ 0.5 \log_2 N+0.9$  & $ 0.57 \log_2 N-0.03$ \\
$S_{N \log N}$ & $ 0.78 \log_2 N-1.15$  & $ 0.91 \log_2 N-2.46$ \\
$S_{0.1 N^2}$ & $0.012N+1.32 \log_2 N$  & $0.004N+2.02 \log_2 N $\\
& $-4.3$ & $-8.47$ \\
\hline
\end{tabular}
\end{center}

We can further probe to which extent the asymptotic behavior for large systems \eqref{eq:boundD} is satisfied within our data.
Without taking  the large $N$ limit in
the sum \eqref{eq:Dmax}, we obtain for arbitrary sizes
\begin{align}
D \lesssim
2 \left[1+g(N) \right] D_0^{ 1/\delta}-\left[1+2 g(N) \right ],
\label{eq:DmaxN}
\end{align}
where we have defined $g(N)=(D_1^{2/\gamma \sqrt{N}}-1)^{-1}$.
In the limit of large system size, $g(N)\sim \gamma \sqrt{N}/2  \log D_1 \gg 1$,
and  \eqref{eq:boundD} is recovered.
We thus estimate $D_0$ from the moderate sizes available by fitting the data for each system size to a function $2^S=a D_0^{1/\delta}+b$.
This yields $D_0^{\mathrm{(Ising)}} \sim 1.50(15)$ for the Ising and $D_0^{\mathrm{(XYZ)}}\sim 3.9(0.24)$ for the XYZ model (where the errors are estimated from the weighted mean of $D_0$ values obtained for each system size $N>20$).
For these values of $D_0$, we check that most of our data satisfies $2^S \propto \sqrt{N} (D_0^{1/\delta}-1)$,
as shown in figure~\ref{fig:cheby_SvsDelta}.
As expected, the largest deviations are observed for the smallest system size $N=20$. For larger systems, the Ising data fits well the
expected behavior, while in the XYZ case the finite size effects seem to be more important, and deviations can be appreciated also for $N=40$.

\begin{figure}[h]
\centering
\subfloat[Ising model]{\label{fig:cheby_Dtr_funN_Ising}\includegraphics[width=.48\columnwidth]{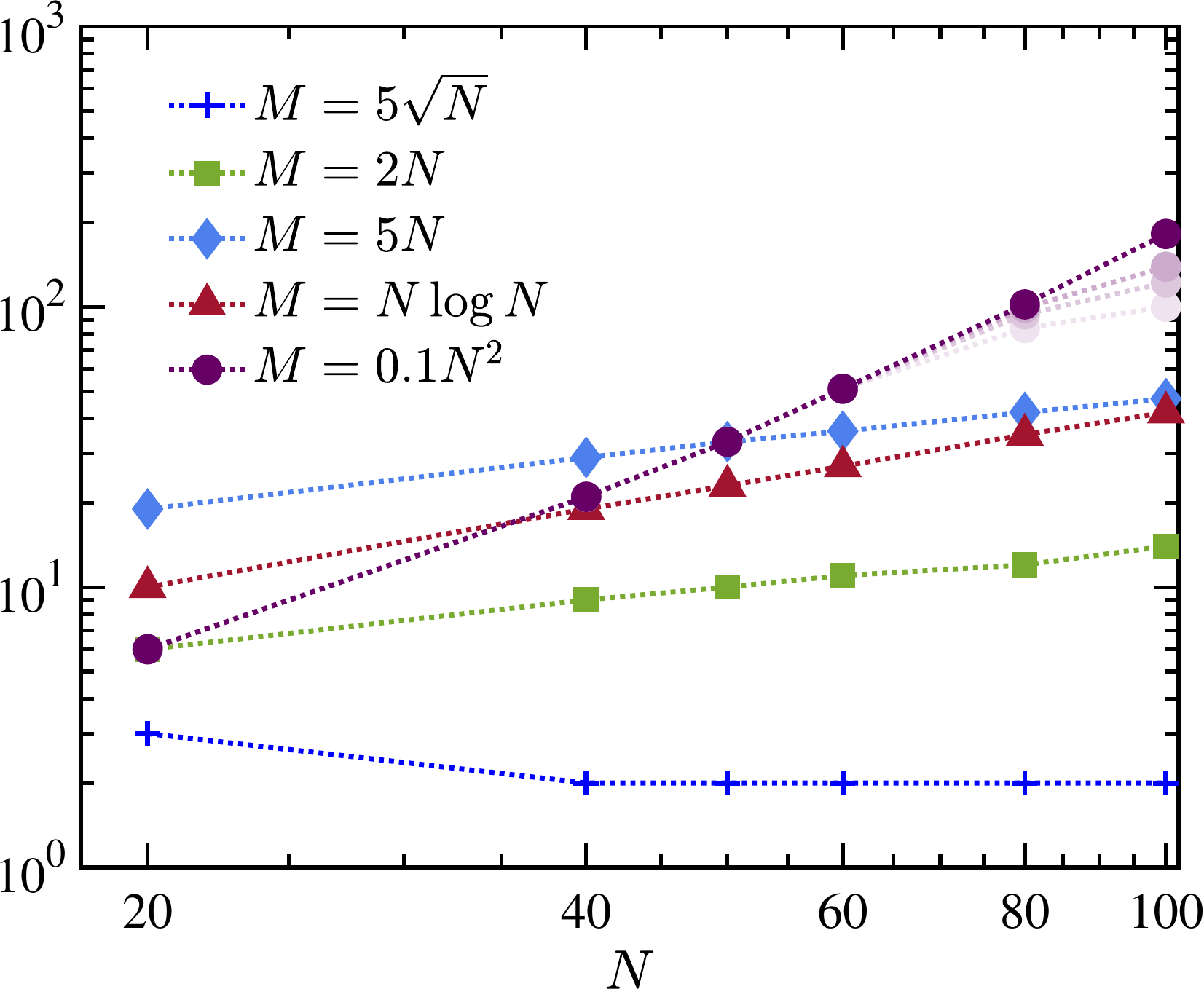}}
\hspace{0.1pt}
\subfloat[XYZ model]{\label{fig:cheby_Dtr_funN_Heis}\includegraphics[width=.48\columnwidth]{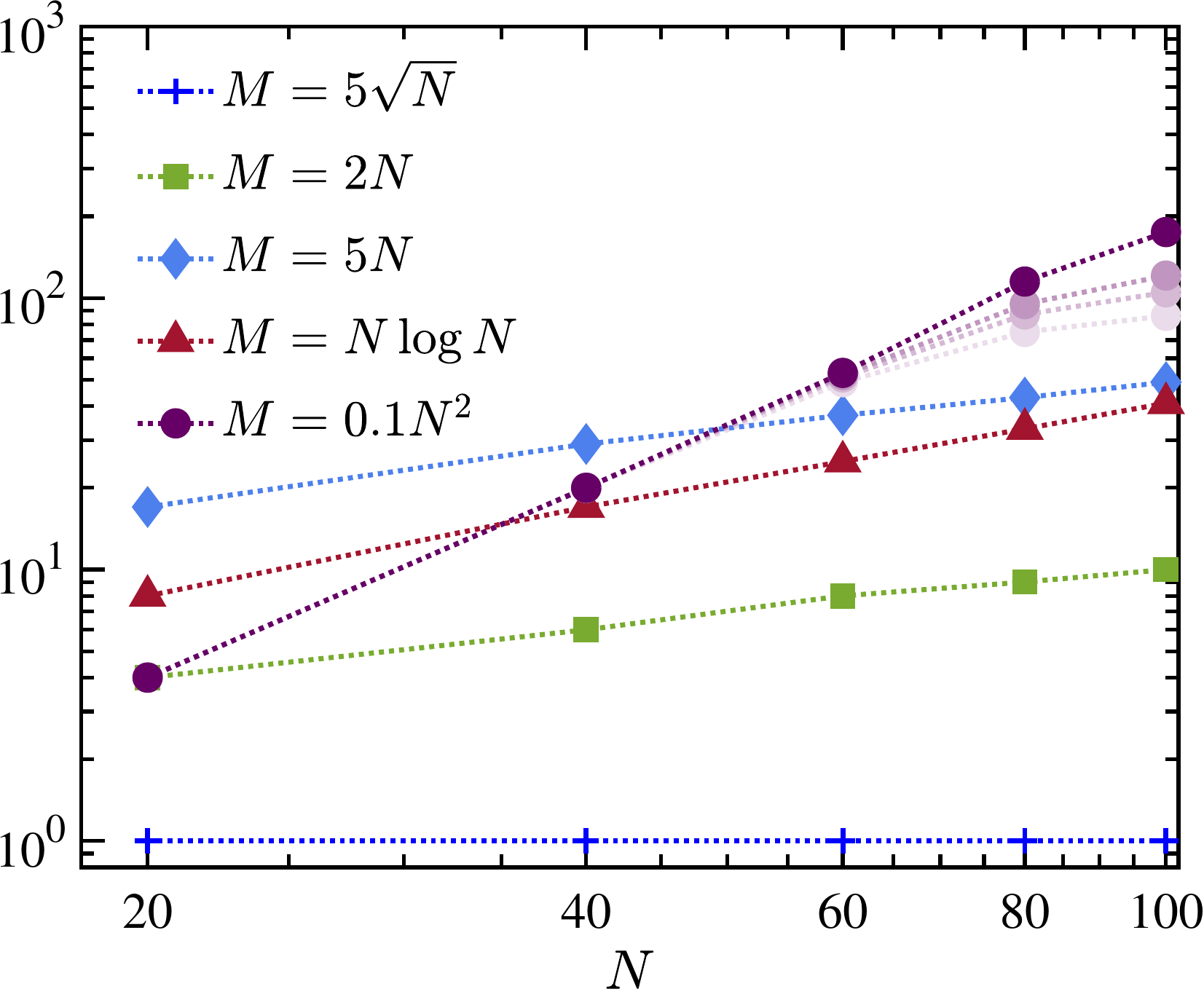}}
\caption{
Minimum bond dimension $\Dtr$ required to keep a truncation error (for the middle cut) smaller than $\epsilon=0.01$
as a function of the system size,
for different truncations of the Chebyshev expansion $M=f(N)$ for the Ising (upper) and XYZ (lower) setups.
As for previous figures, solid symbols correspond to $D=1000$.
Darker colors correspond to larger bond dimensions.
We observe that almost all cases grow polynomially with $N$,
except for $M=0.1 N^2$, when the values are also compatible with an additional linear term in $N$.
}
\label{fig:cheby_Dtr_funN}
\end{figure}

From our simulations, run with constant bond dimension, we can estimate the truncation error and thus the minimal bond dimension
required to maintain a given precision in the MPS representation of the states we construct.
The MPS form gives access to the Schmidt decomposition $\{\lambda_k\}_{k=1}^D$ across any cut.
For the  one corresponding to the middle of the chain
 we define $\Dtr$ as the minimum bond dimension required to ensure a small error $\epsilon=10^{-2}$,
namely $1-\sum_{i=1}^{\Dtr}|\lambda_k|^2\leq\epsilon$.
We then analyze the scaling of $\Dtr$ as we did for the entropy.
Figure~\ref{fig:cheby_Dtr_funN} shows the scaling with the system size for the various families of states $M=f(N)$. We observe that almost all cases are compatible with a polynomial
increase $\Dtr \propto N^{\beta}$, with $\beta=0,$ $0.42,$ $0.53$ and $0.88$, respectively,
for $M=5\sqrt{N},$ $2N,$ $5N$ and $N \log N$ in the Ising case and
$\beta=0,$ $0.59,$ $0.57$ and $0.96$ in the XYZ model.
The case $M=0.1 N^2$, instead, shows a faster increase and is compatible with a fit $\log_2 \Dtr\sim 0.02 N + 1.47 \log_2 N-7.9$ (Ising)
and $\log_2 \Dtr\sim 0.02 N + 1.70 \log_2 N-5.5$ (XYZ). Notice that these estimates are obtained for the largest $D=1000$ (discarding the largest sizes which may not be fully converged).

Repeating the analysis we described for the entropy, we can fit the data for each system size to a curve $\Dtr=a D_0^{1/\delta} + b$,
extract $D_0$ and check the scaling \eqref{eq:boundD}. Again we observe
 $\Dtr/\sqrt{N}$ varying linearly in $D_0^{1/\delta}$ (see insets of figure~\ref{fig:cheby_SvsDelta}).

\subsection{Local similarity to thermal state}
\label{subsec:local}

\begin{figure}[h]
\centering
\subfloat[Ising model, trace distance $L_c=8$]{\label{fig:cheby_dist_funN_Lc1_Ising}\includegraphics[width=.5\columnwidth]{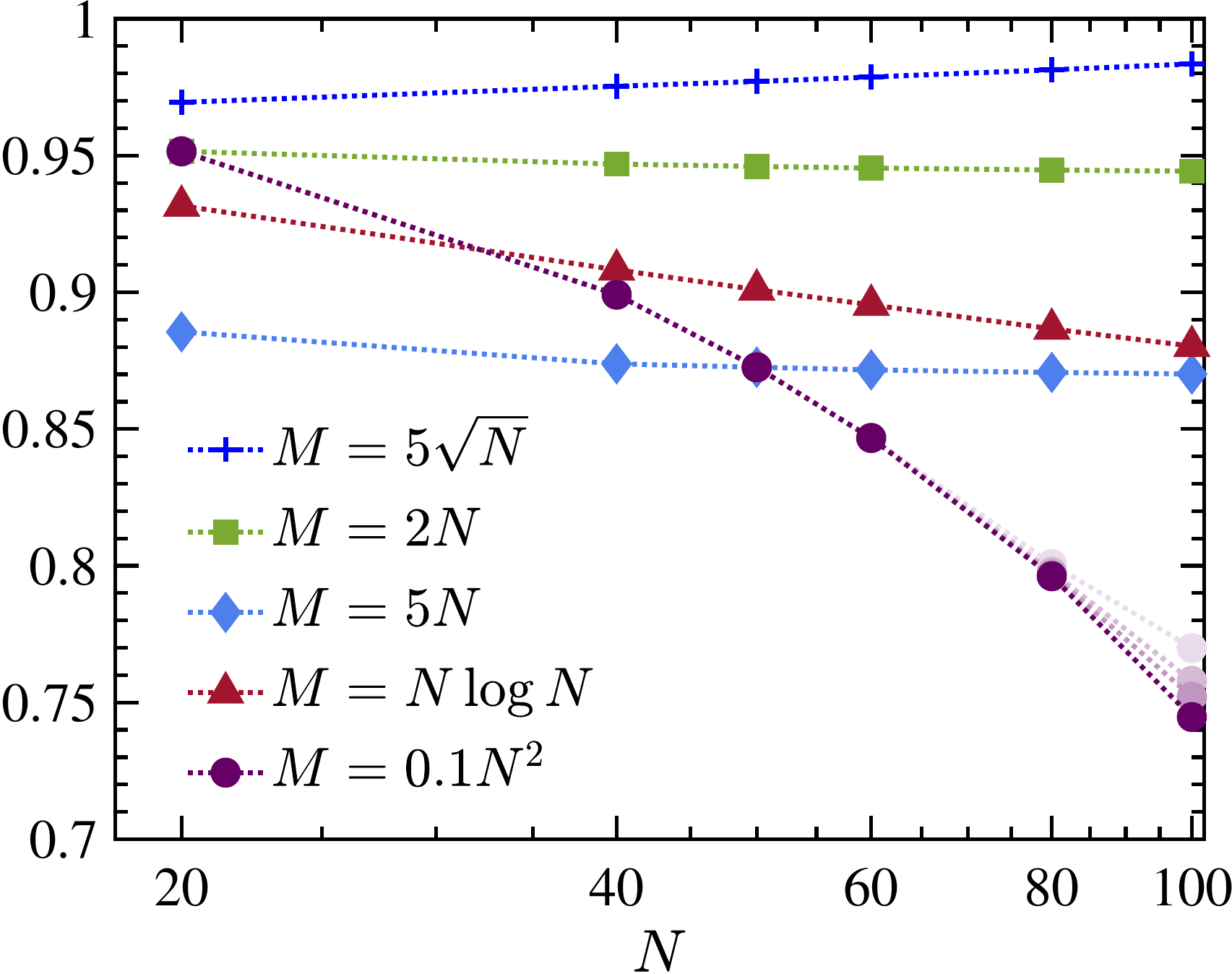}}
\hspace{.1pt}
\subfloat[Ising model, entropy $L_c=8$]{\label{fig:cheby_dist_funN_Lc2_Ising}\includegraphics[width=.47\columnwidth]{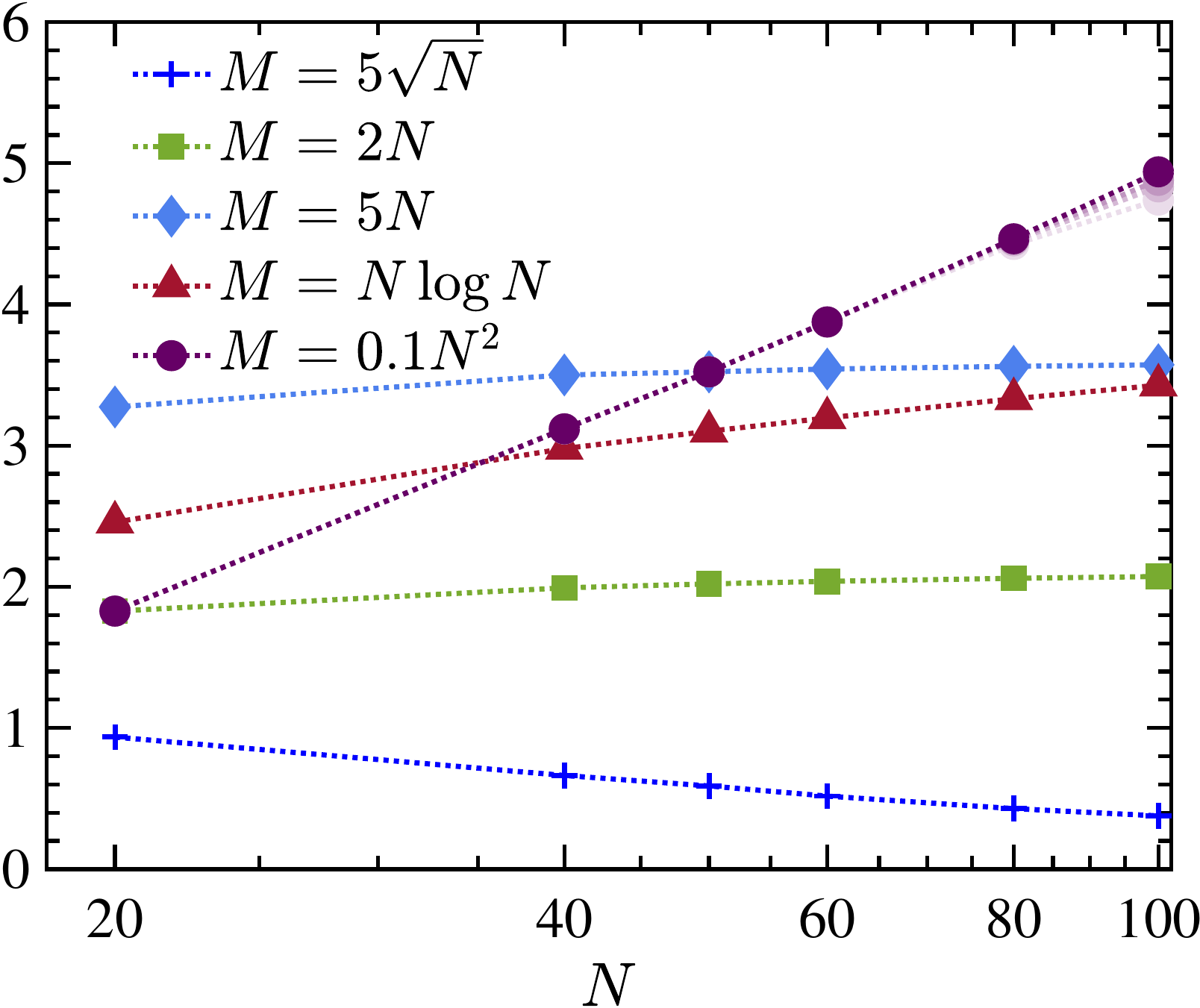}}\\
\subfloat[XYZ model,  trace distance $L_c=8$]{\label{fig:cheby_dist_funN_Lc1_Heis}\includegraphics[width=.5\columnwidth]{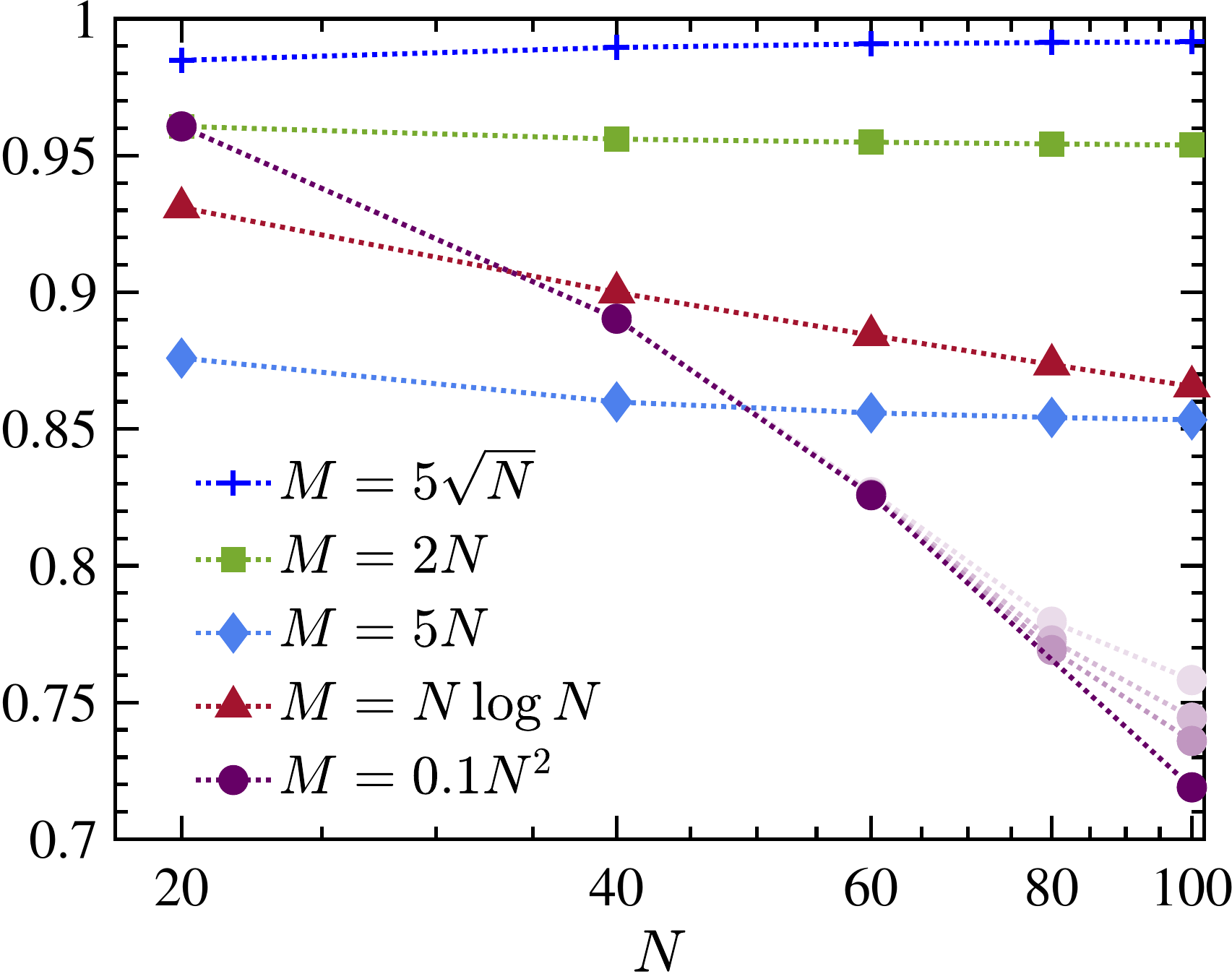}}
\hspace{.1pt}
\subfloat[XYZ model, entropy $L_c=8$]{\label{fig:cheby_dist_funN_Lc2_Heis}\includegraphics[width=.47\columnwidth]{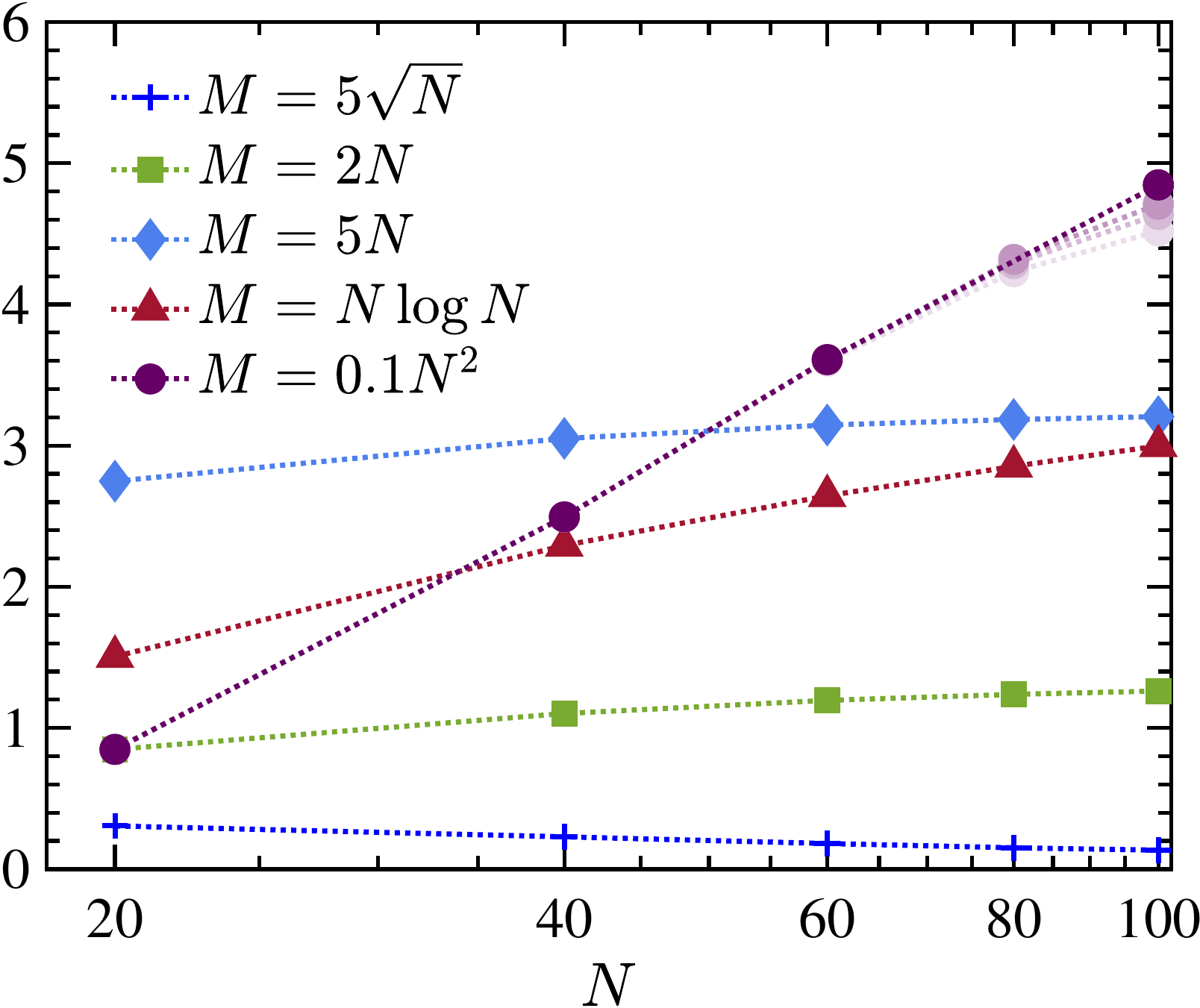}}
\caption{
Trace distance (left column) between the reduced density matrix for the $L_c=8$ central sites and the thermal state at the end of the Chebyshev filter,
corresponding to data in figure~\ref{fig:cheby_Delta2vsN}.
For $M\propto\sqrt{N}$, which gives approximately $\delta \propto \sqrt{N}$, the distance grows and approaches the maximum ($0.996$).
For constant variance, instead, the distance seems to stay almost constant in the Ising case, and to go slightly down in the XYZ case.
A number of steps $M\propto N \log N$ or $M \propto N^2$ seems to be enough to have convergence of the
local reduced density matrices to the thermal one.
As in previous figures, solid symbols correspond to the maximum bond dimension $D=1000$, and for the largest number of steps we show in lighter shades the non-converged results for $D=300-500$.
The right column shows the entropy of the corresponding reduced density matrix for the same cases.
}
\label{fig:cheby_dist_funN_Lc8}
\end{figure}

For all the states explored in  figure~\ref{fig:cheby_Delta2vsN_Ising},
i.e. sets of states defined by a truncation order $M=f(N)\geq \sqrt{N}$,
the scaling of the variance is compatible with $\delta\sim N^{\eta}$, with $\eta<1$,
so that the energy density variance $\delta/N\sim N^{\eta-1}$  vanishes in the thermodynamic limit.
This is however not enough to guarantee that the reduced state for a subsystem is close to thermal.

In the setups we consider, the (target) mean energy is zero, which corresponds to infinite temperature.
We thus compare the reduced density matrix for the central $L_c$ sites (up to $L_c=10$)
with the corresponding reduced density matrix for the maximally mixed state.
We study how this distance varies as a function of the system size
for the different sets of states described above, with
different scalings of the variance.
The results for $L_c=8$ are shown in figure~\ref{fig:cheby_dist_funN_Lc8}.

We observe that for $M\propto \sqrt{N}$, which yields $\delta\sim \sqrt{N}$, the local distance for all sizes $L_c$
grows with the system size, approaching the maximum possible value, $1-2^{-L_c}$.
For linear truncation $M\propto N$, which results into almost constant (or very slightly decreasing) $\delta$,
the local distance seems to become constant, growing a bit for the smallest systems in the Ising case, and decreasing in the XYZ case, but seemingly
stabilizing for larger sizes. These states, thus, do not locally resemble thermal equilibrium either in the thermodynamic limit.
Our results suggest however that an energy variance decreasing as $\log N$ or faster
would guarantee local convergence to thermal equilibrium in the limit of large $N$
(as the curves for $M=N \log N$ and $M=0.1 N^2$ illustrate).
We observe the same qualitative behavior in both setups. Also different subsystem sizes (except the smallest ones) behave similarly.

A necessary condition for the reduced state of $L_c$ sites to resemble thermal equilibrium (which here corresponds to the
maximally mixed state) is that the entropy of the corresponding subsystem
grows as $L_c$.
We have evaluated the entropy of the central subchain in these states for $L_c=1,\ldots 10$.
Our arguments bound the growth of this entropy with the same form~\eqref{eq:boundS},
but since the maximum entropy of the block is $L_c$, if $L_c < \log N /2$,
 the right hand side of~\eqref{eq:boundS} just gives a trivial bound.
 Thus we illustrate the behavior for the case $L_c=8$ in the right panels of fig.~\ref{fig:cheby_dist_funN_Lc8},
 although qualitatively similar plots are obtained for the other sizes explored, if $L_c\gtrsim 4$.
We observe that, while for large entropy of the block, approaching the upper bound $L_c$,
the distance decreases much faster than $L_c-S(\rho_{L_c})$,
when the entropy is comparatively smaller, there is a clear correlation between both quantities,
and the entropy fulfills the form~\eqref{eq:boundS} qualitatively.

\subsection{Variance and correlations}
\label{subsec:corr}

\begin{figure}[h]
\centering
\subfloat[Ising model]{\label{fig:Ecorr_Ising}\includegraphics[width=.48\columnwidth]{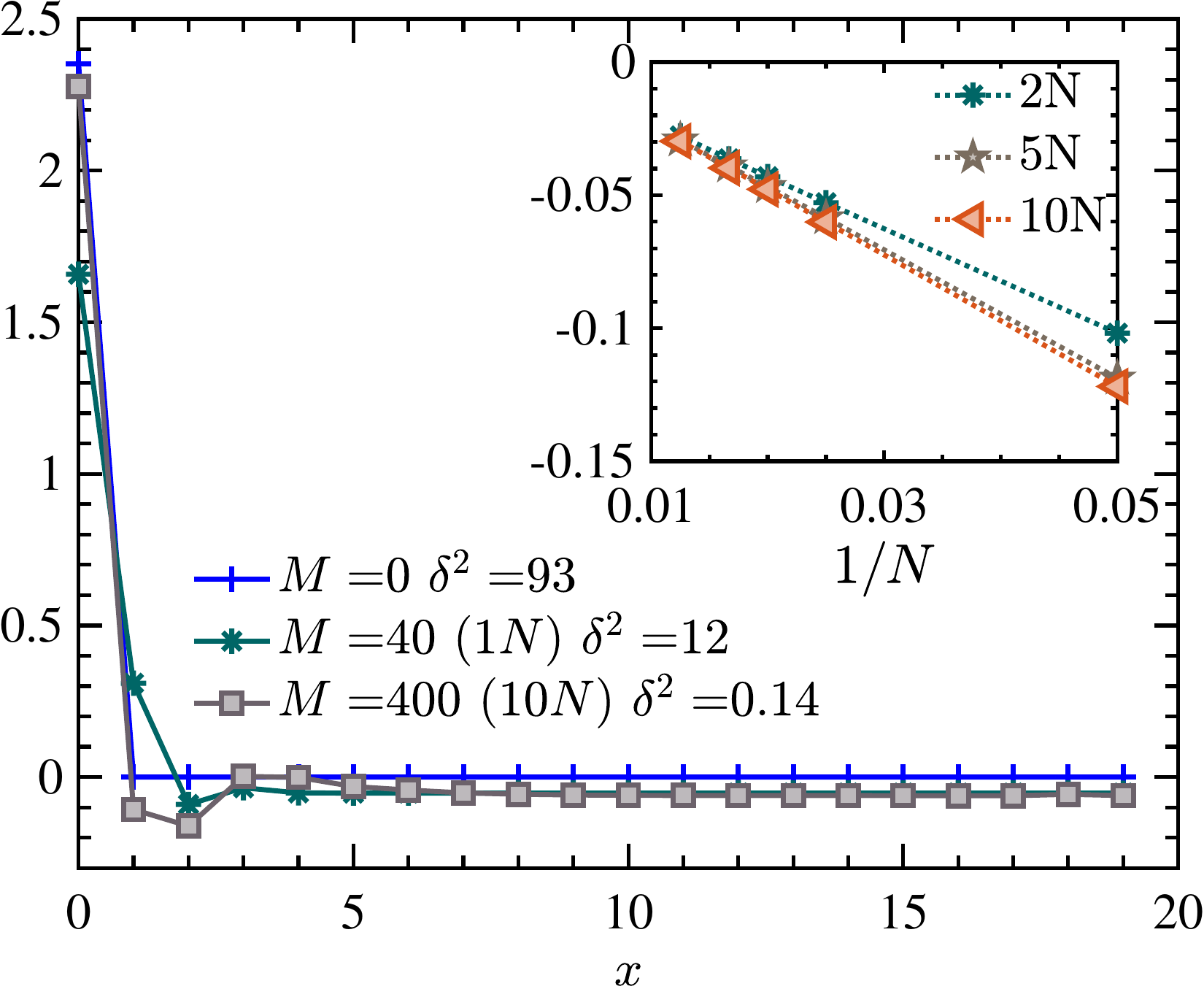}}
\hspace{.1pt}
\subfloat[XYZ model]{\label{fig:Ecorr_XYZ}\includegraphics[width=.48\columnwidth]{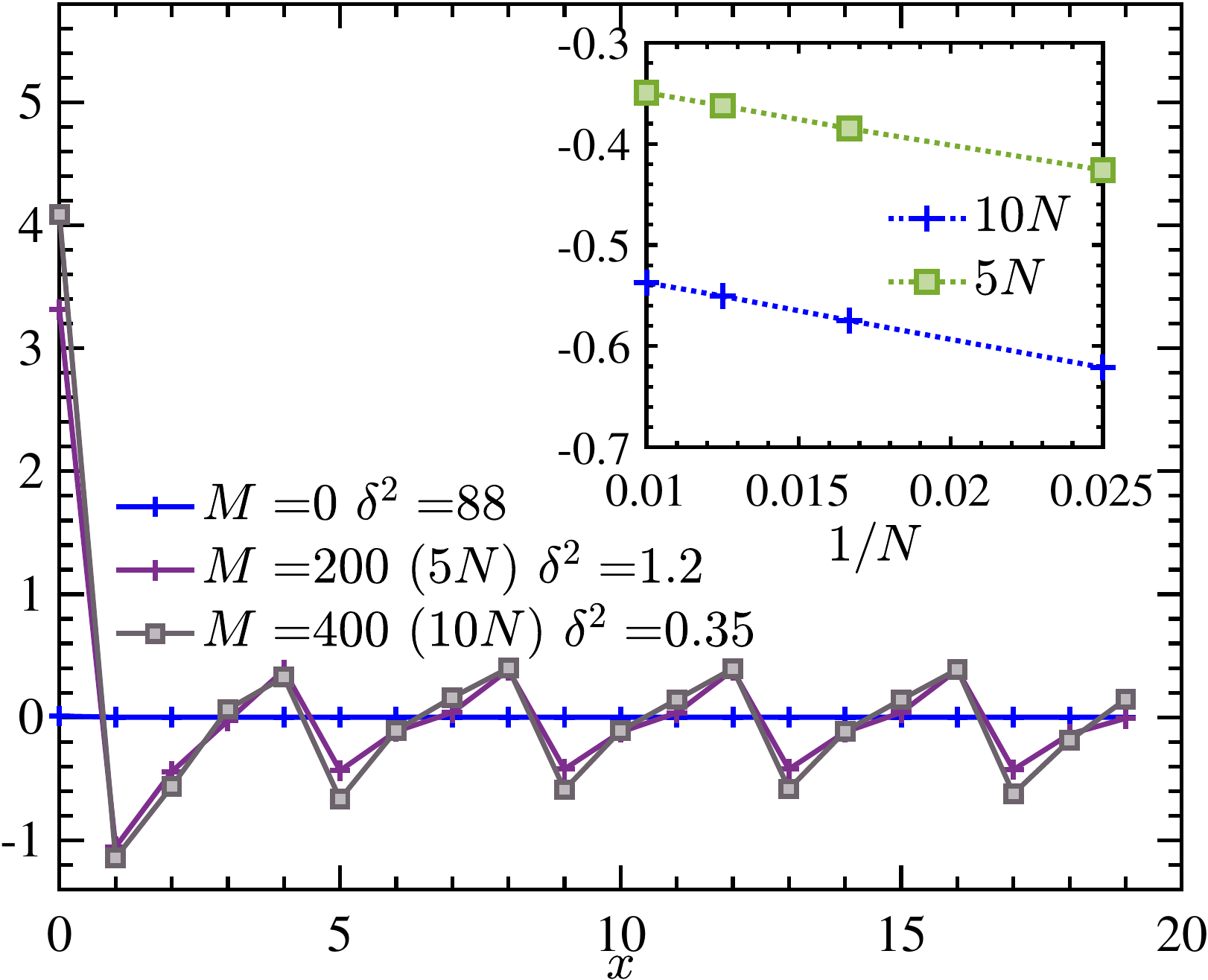}}
\caption{
Spatial distributions of energy density correlations $\ev{h_{n_c}h_{n_c+x}}-\ev{h_{n_c}}\ev{h_{n_c+x}}$ with respect to a site $n_c$ near the center of the chain\cite{Note3} ($N=40$) in both models,
for different number of steps $M$.
The insets show the approximate $1/N$ dependence of long range terms (magnitude of the peaks for XYZ) for numbers of iterations $M\propto N$ that keep the variance constant.
}
\label{fig:corrE}
\end{figure}

In order to decrease the energy variance $\delta^2=\sum_{n,m}\bra{\Psi} h_n h_m \ket{\Psi}-\bra{\Psi}h_n|\Psi\rangle \bra{\Psi}h_m|\Psi\rangle$,
the system needs to arrange the local energy fluctuations and their correlations, such that the sum nearly vanishes.
For a translationally invariant system with zero mean energy, and taking into account that $\ev{h_n^2}\geq a$ for some constant $a>0$,
a constant $\delta^2$ can thus be attained by either some short range terms $\ev{h_n h_{n+\ell}}$ of $O(1)$, or if all terms $\ev{h_n h_m}$ become $O(1/N)$.
By inspecting the spatial distribution of such correlators in the states constructed in this section,
 we can thus better understand how our method constructs the states with small energy variance (the local energy operators $h_n$ are
 chosen to fulfill $\mathrm{tr}_n h_n=0$, as specified in section~ \ref{sec:prelim}).

As illustrated in figure~\ref{fig:corrE} for the case $N=40$, very early a certain amount of long range correlations develops. In the Ising case (left panel),
where the initial state is translationally invariant, these correlations are homogeneous, while in the XYZ model (right panel) they reflect the periodicity of the initial state.
\footnote{The reference site is chosen to be the central one  $n_c=N/2$ in the Ising case. In the non-translationally invariant XYZ case, in order to compare different system sizes 
we choose as reference the position of the $\ket{0011}$ substring closest to the center in the initial state.}
By comparing the (largest) magnitude of the long range correlations for different system sizes, at a number of steps $M\propto N$, which, as we have
discussed, corresponds to a constant $\delta^2$, we observe that these long-range terms scale indeed as $1/N$.
This strategy is thus not the most entanglement effective, which explains why, when we search for the states minimizing the variance at fixed energy and bond
dimension (see fig.~ \ref{fig:var_corrE}), we encounter different structure.

\section{Inhomogeneous energy distribution}
\label{sec:inhom}

\begin{figure}[h]
\centering
\includegraphics[width=.8\columnwidth]{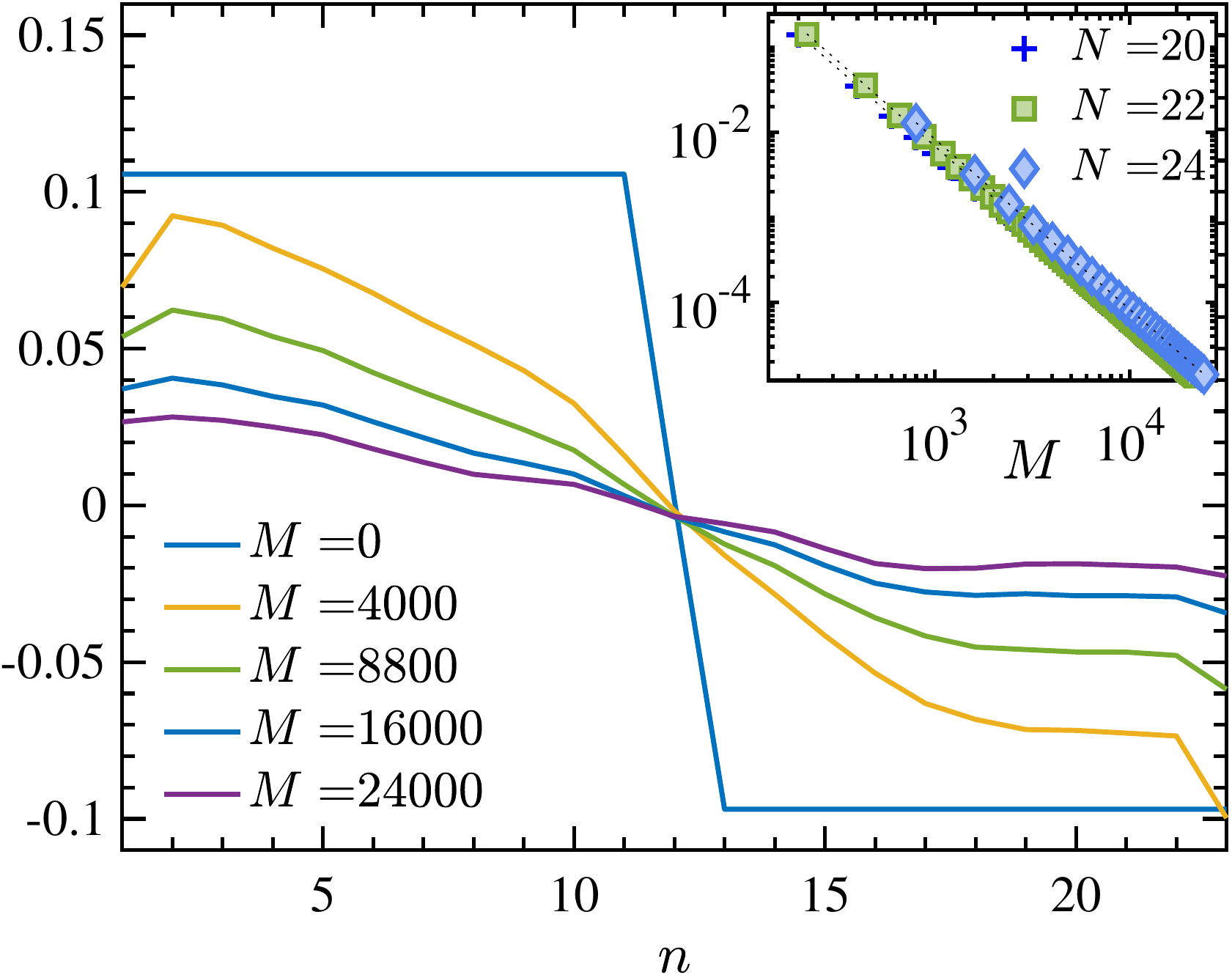}
\caption{
Applying the Chebyshev filter to initial states with inhomogeneous energy density. The variance (inset) decreases as $\delta^2\propto 1/M^2$: the dotted black lines
that fit the data to $\delta^2 =A M^{-\eta}$ give precisely exponent $\eta=2$.
The main panel shows the local energy density as a function of the position in the chain (for $N=24$) for some values of $M$.
The initial distribution (bright blue) is a step, and the filter smears it slowly: after $M=8800$ steps the inhomogeneity is still significant (green), while the variance read in the inset
has decreased by three orders of magnitude until $\delta^2\sim 10^{-4}$.
}
\label{fig:stepE}
\end{figure}

The scaling of the variance in our constructions \eqref{eq:deltaCos} and \eqref{eq:deltaCheby}
follows from considering an initial product state with narrow enough ($\sigma_p \sim \sqrt{N}$) energy distribution,
but does not require translational invariance, and thus
must also hold for initial states that have an inhomogeneous energy density.
Simulating this situation allows us to
ask how such an initial imbalance diminishes as the variance is reduced.

To probe this scenario numerically we consider a product initial state with
a  step-wise energy density, and zero mean energy, with respect to the non-integrable Ising Hamiltonian,
and apply the Chebyshev filter as described above.
Figure~\ref{fig:stepE} show the results of our simulation, using exact diagonalization, for systems up to $N=24$ sites
(although we examined larger systems with MPS, the truncation error became important much before the
effects on the spatial profile are noticeable).
We observe in the inset that the scaling of the variance decreases with the number of steps
as $\delta^2\propto 1/M^2$, as expected from~\eqref{eq:deltaCheby}.
The inhomogeneity of the energy density, nevertheless, survives much longer, and it
remains noticeable even when the variance has decreased by several
orders of magnitude.

\section{Variational optimization}
\label{sec:variational}

\begin{figure}[h]
\centering
\subfloat[Ising model]{\label{fig:var_DeltaVsD_Ising}\includegraphics[width=.48\columnwidth]{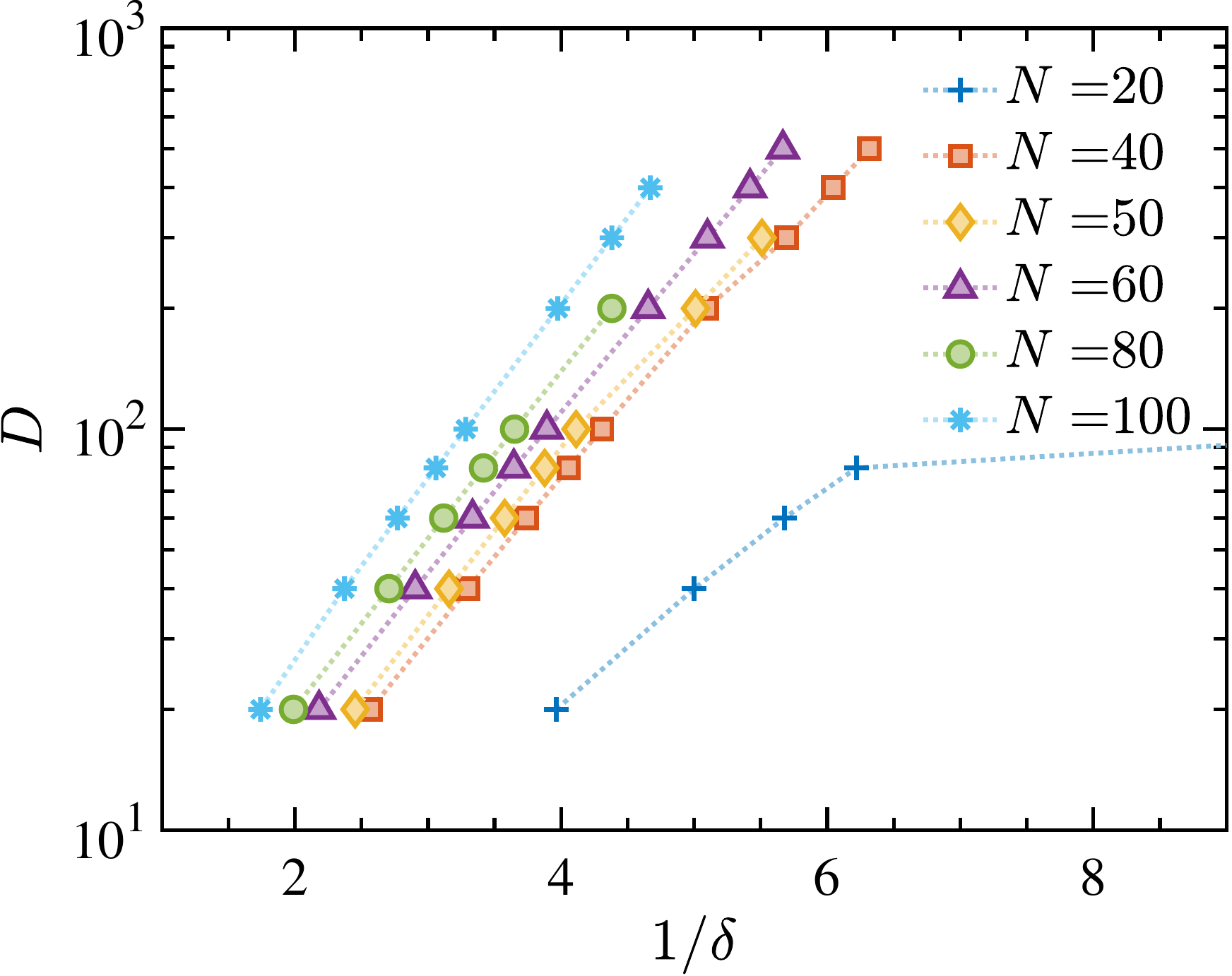}}
\hspace{0.1pt}
\subfloat[XYZ model]{\label{fig:var_DeltaVsD_Heis}\includegraphics[width=.48\columnwidth]{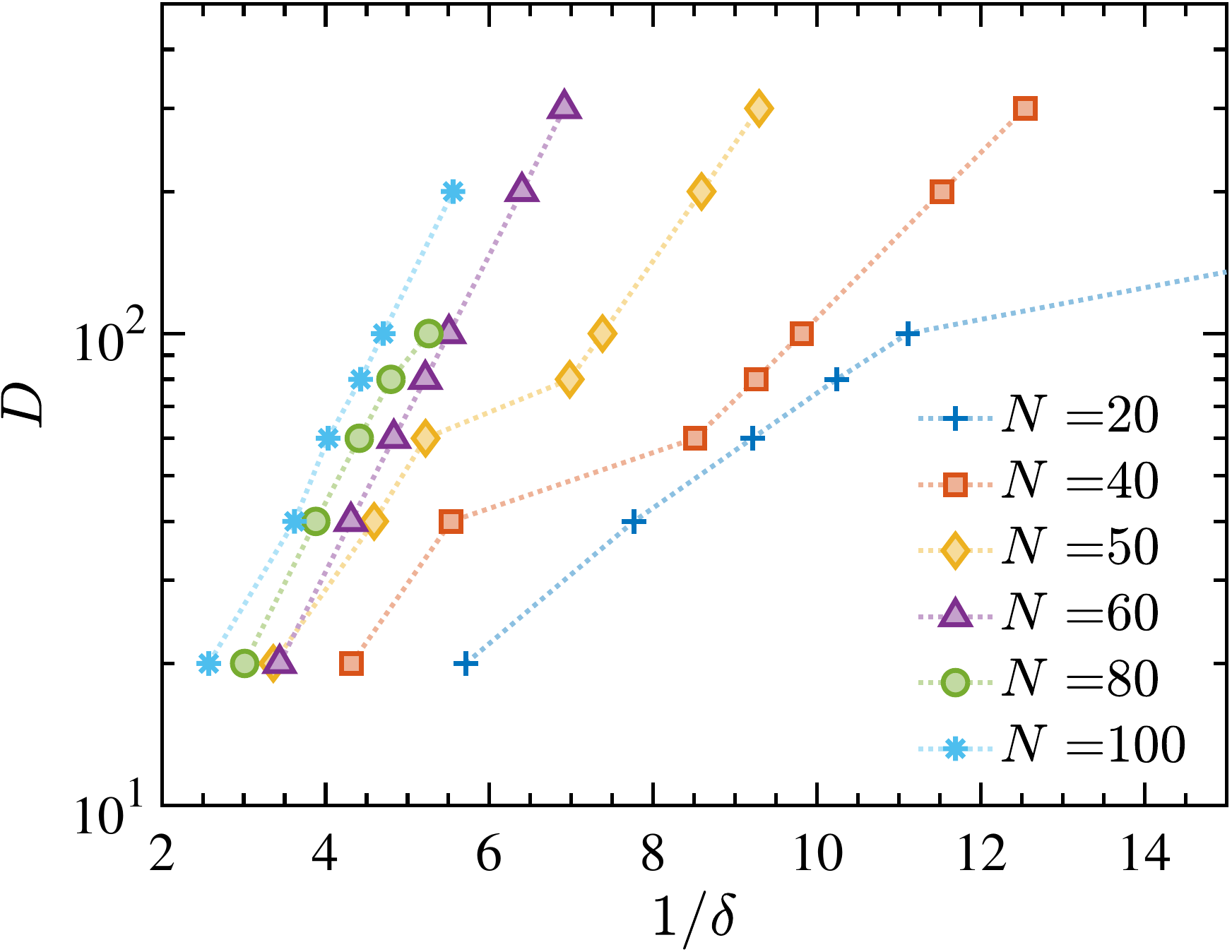}}
\caption{Relation between the bond dimension and the final energy variance attained via variational minimization
for several system sizes for the Ising (above) and XYZ (below) setups.
For large enough system sizes we observed a behavior compatible with $\log D \sim 1/\delta$.} 
\label{fig:var_DeltaVsD}
\end{figure}

\begin{figure}[h]
\centering
\subfloat[Ising model]{\label{fig:var_DistVsD_Ising}\includegraphics[width=.49\columnwidth]{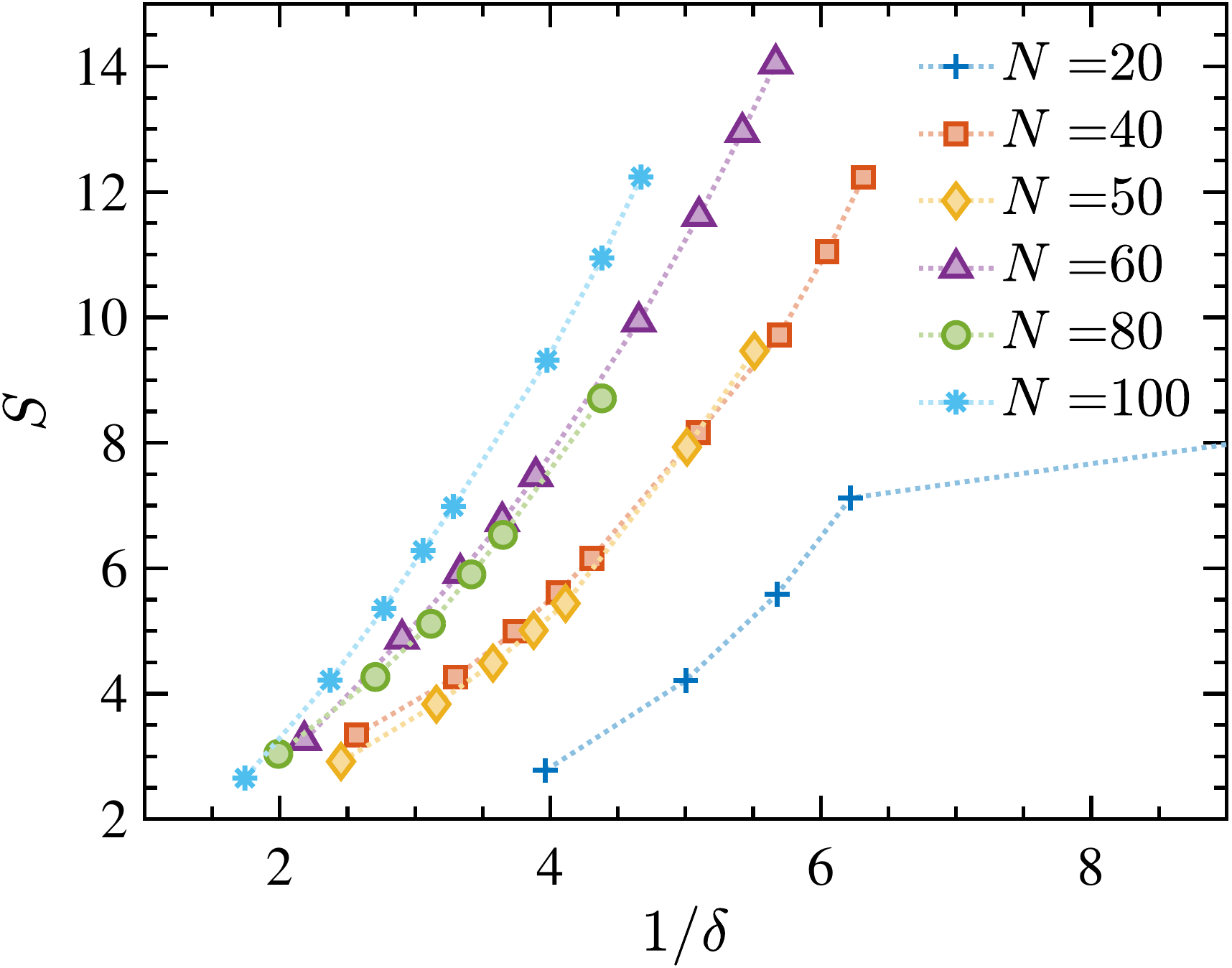}}
\hspace{0.1pt}
\subfloat[XYZ model]{\label{fig:var_DistVsD_Heis}\includegraphics[width=.48\columnwidth]{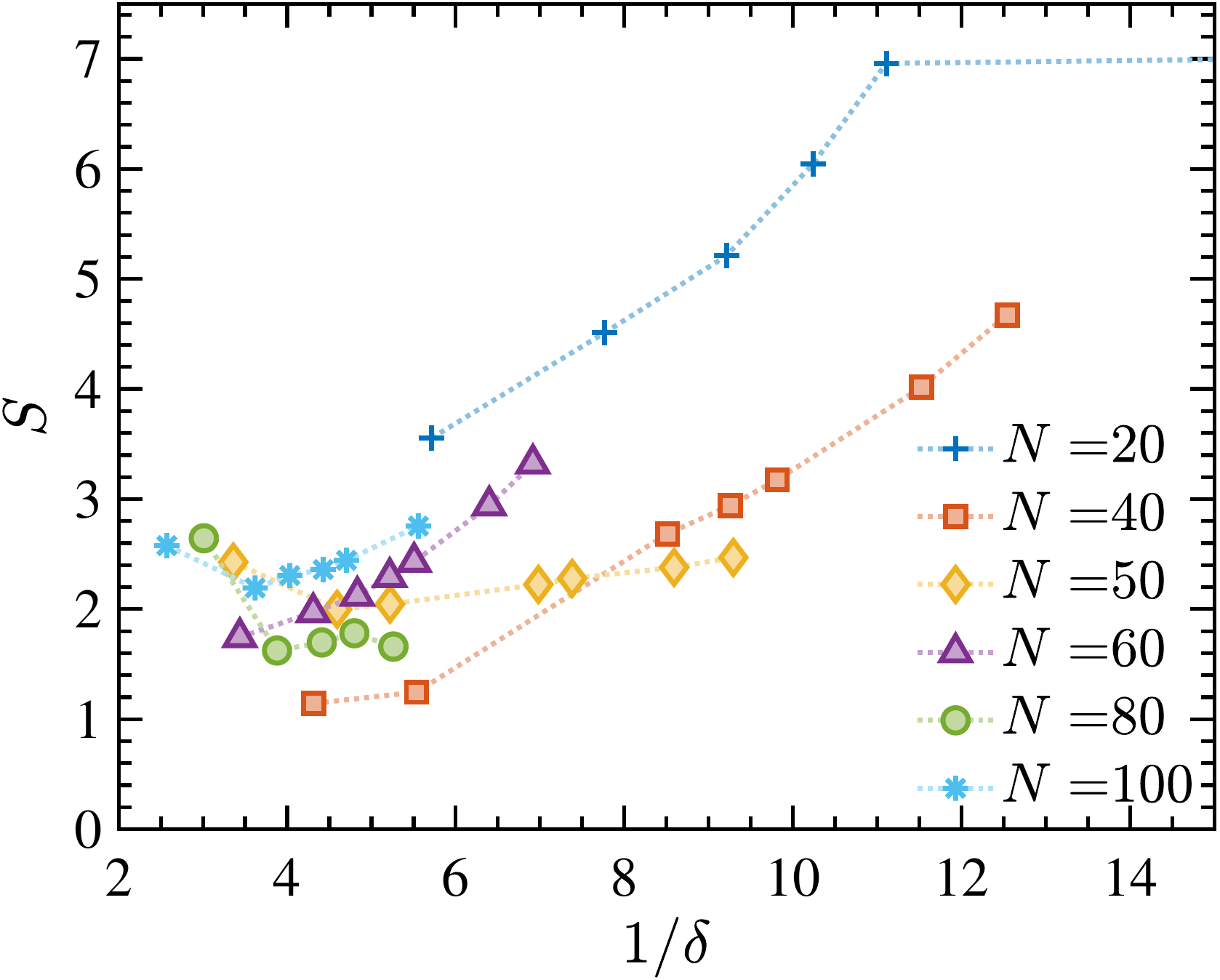}}
\caption{
Entanglement entropy (half chain) of the MPS resulting from the variational minimization of the variance, as a function of  $1/\delta$,
for different system sizes for the Ising (above) and XYZ (below) setups. }
\label{fig:var_SVsD}
\end{figure}

\begin{figure}[h]
\centering
\subfloat[Ising model]{\label{fig:var_DistVsD_Ising}\includegraphics[width=.48\columnwidth]{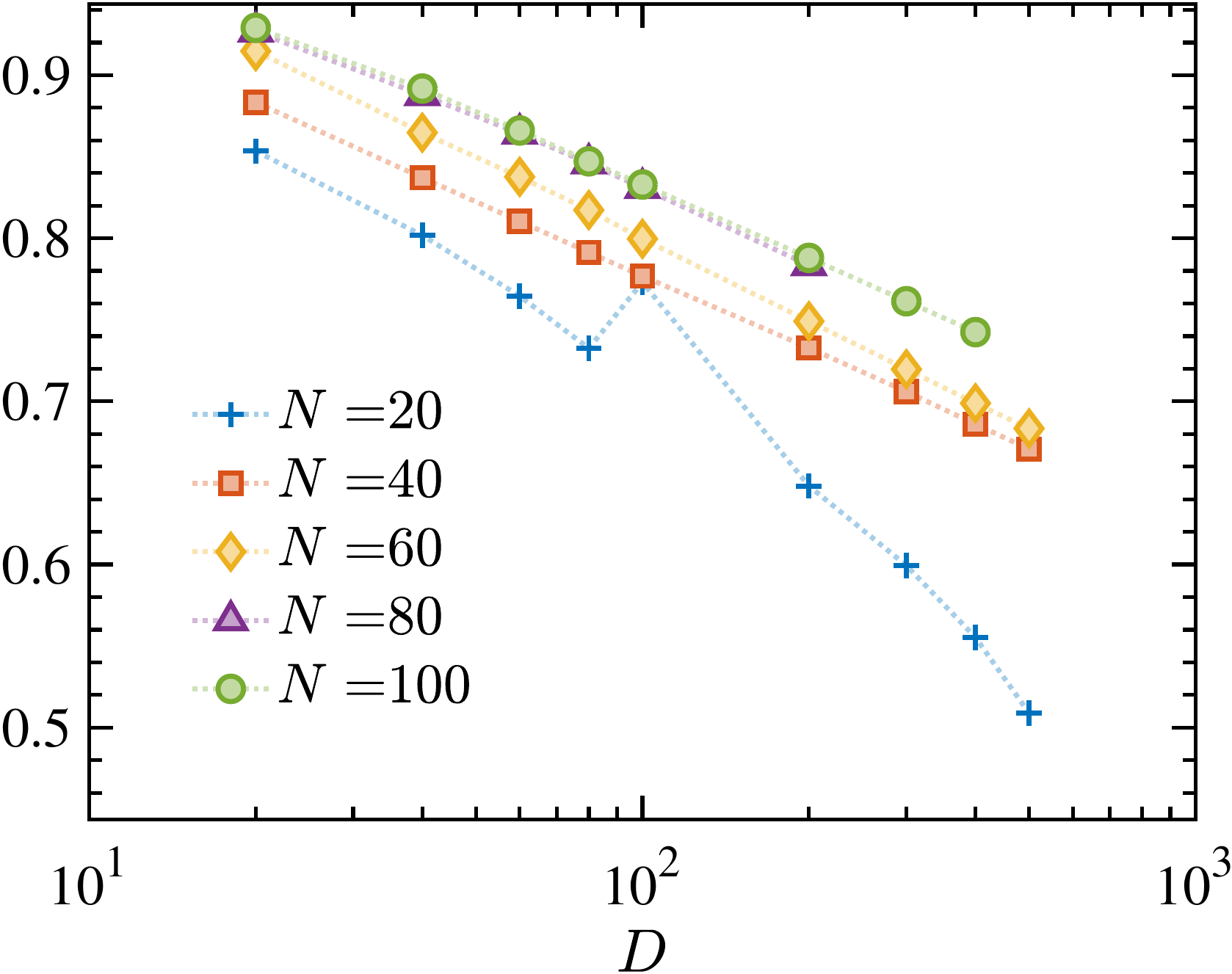}}
\hspace{0.1pt}
\subfloat[XYZ model]{\label{fig:var_DistVsD_Heis}\includegraphics[width=.49\columnwidth]{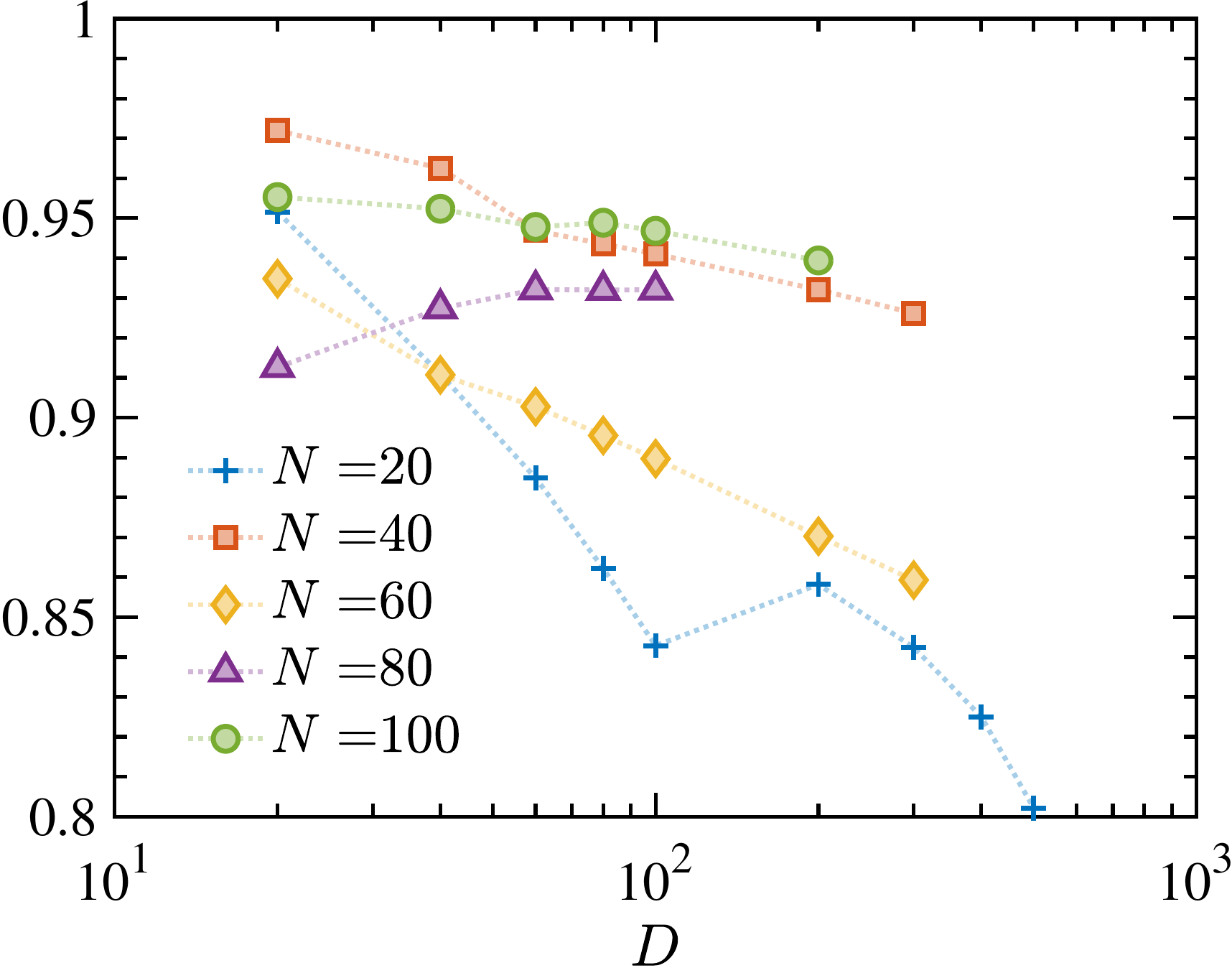}}
\caption{
Average trace distance between a $L_c=8$ subsystem and the thermal equilibrium state at infinite temperature, as a function of the bond dimension,
after the variational minimization, for several system sizes for the Ising (left) and XYZ (right) setups. The lines are simply for visual aid. }
\label{fig:var_DistVsD}
\end{figure}

\begin{figure}[h]
\centering
\subfloat[Ising model]{\label{fig:var_Ecorr_Ising}\includegraphics[width=.48\columnwidth]{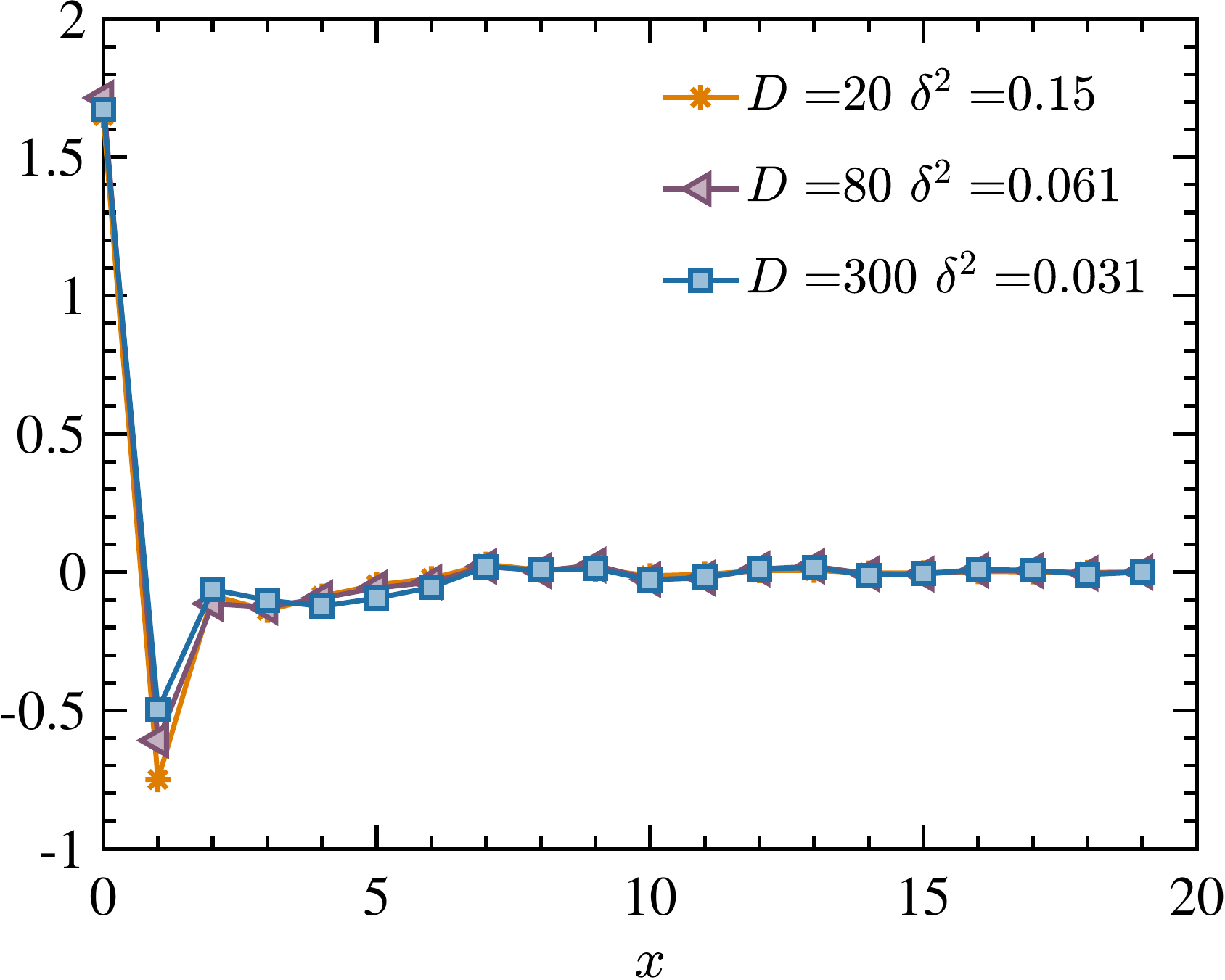}}
\hspace{.1pt}
\subfloat[XYZ model]{\label{fig:var_Ecorr_XYZ}\includegraphics[width=.48\columnwidth]{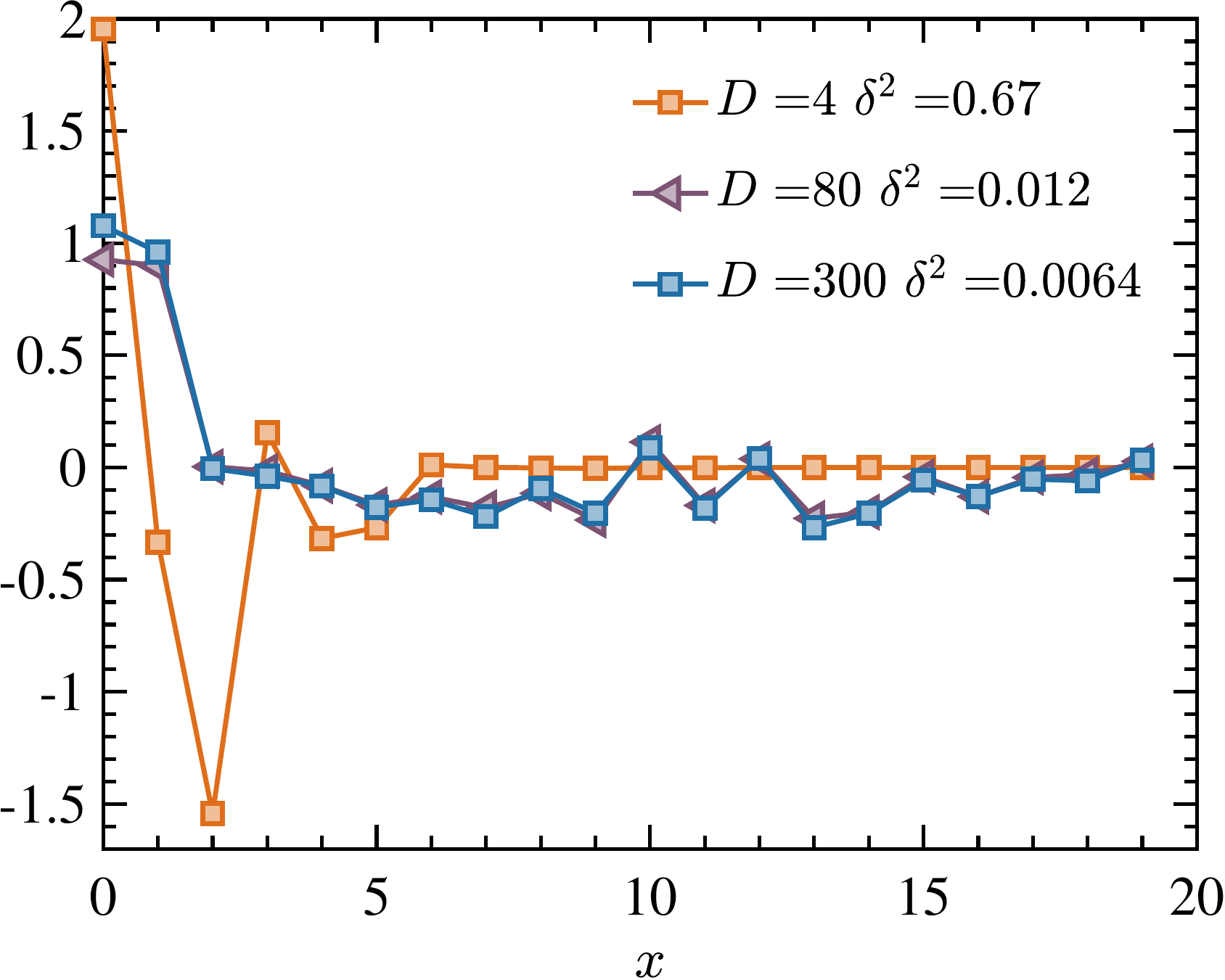}}
\caption{
Spatial distributions of energy density correlations $\ev{h_{n_c}h_{n_c+x}}-\ev{h_{n_c}}\ev{h_{n_c+x}}$ with respect to a central site $n_c$ of the chain ($N=40$) in both models,
for the states minimizing the variance at fixed $D$ (to be compared to figure~\ref{fig:corrE}).
}
\label{fig:var_corrE}
\end{figure}

Applying the strategy described in the paper with a fixed bond dimension manages to
produce a MPS with small energy variance at the given mean energy.
This variance can be systematically reduced by increasing the order of the
Chebyshev expansion, before truncation error appears.
But this does not need to be the MPS with the smallest possible variance
for the given bond dimension and mean energy.
Instead, we can directly search for such optimal MPS by a variational optimization using as cost function
$\Delta H^2=\bra{\Psi}H^2\ket{\Psi}- \bra{\Psi}H\ket{\Psi}^2$ (plus a penalty term to ensure the desired mean energy).
This variational problem can be formulated as the optimization of a MPS,
and be solved using a sequential iteration over tensors, similar to DMRG algorithms~\cite{verstraete08algo,schollwoeck11age},
with the difference that the local problem to be solved for each individual tensor is the
minimization of a quartic expression. A similar problem appears also when optimizing purifications and can
be solved using some iterative numerical scheme, e.g. a gradient descent algorithm (see e.g.~\cite{lubasch2014peps}),
but the problem has local minima (even at the level of the individual tensors) and the
convergence severely depends on the parameters of this local optimizer.

We have performed the variational search for the same setups and system sizes discussed in the first part of the paper,
and using bond dimensions $20\leq D\leq 500$,
in order to compare the results to the ones discussed in the previous section.
We observe that the values of the energy variance reached for a certain bond dimension
can be much smaller than with the Chebyshev sum.
Similar to that case, the bond dimension seems to grow exponentially with $1/\delta$,
with a coefficient that depends on the system size,
as shown in figure~\ref{fig:var_DeltaVsD}.
but which does not correspond to $\sqrt{N}$.
From our data we could not identify a clear scaling
of the coefficients.
Additionally, the behavior of the variance is not smooth when increasing $D$,
what we attribute to the imperfect convergence of the non-linear optimizations, which
may sometimes be trapped in local minima for a certain value of the bond dimension,
while the state may change completely when the bond dimension is varied.

The entanglement entropy of the states found in the optimization reflects an even stronger non-systematic behavior,
specially in the case of the XYZ model,
as shown in figure~\ref{fig:var_SVsD}.
The average distance of the subsystems to the thermal equilibrium does not behave monotonically either (see figure~ \ref{fig:var_DistVsD}),
with the large changes corresponding to the abrupt variations in $\delta$ appreciated in figure~\ref{fig:var_DeltaVsD}.
Interestingly, although the variance decreases monotonically when increasing $D$, the distance does not behave in the same way.
Overall,
the distances obtained with the variational minima are (except for the smallest system) larger than the
best distances achieved with the filter. 

Also the energy density correlations in these states differ from the systematic behavior encountered in figure~ \ref{fig:corrE}.
Now we find that the states achieve much lower energy variance by modifying the short range correlations, and
long range ones are developed only at much larger bond dimensions (see figure~\ref{fig:var_corrE}).

\section{Discussion}
\label{sec:discussion}

We have introduced a method that,
starting from a product state,
systematically constructs states with decreasing energy variance
and controlled entanglement
for any local one-dimensional Hamiltonian.
This allows us to extract conclusions about
the minimal entanglement (or bond dimension)  guaranteeing
the existence of a state with a certain variance, as a function of the system size.
We have found that it is possible to prepare states with arbitrarily small variance (vanishing as $\delta\sim 1/\log N$)
with a bond dimension that scales polynomially with the system size.

Using MPS algorithms, we have implemented the construction numerically for the Ising and XYZ models,
and we have confirmed that the asymptotic scalings
hold
already for system sizes $N\in\{20,100\}$.
Using the numerical simulation we have also analyzed how close these states are to thermal equilibrium, in terms of the local reduced density matrices.
Our results suggest that a variance that vanishes as $\delta \sim 1/\log N$ is enough to obtain local thermal behavior
for small subsystems.

For comparison, we have run a variational search for the MPS that minimizes the energy variance at fixed bond dimension.
Although the variational method we use may find convergence problems, the states we find
enable a qualitative comparison with the results from the systematic construction.
The variational search finds states with smaller variance for the same entanglement, although in general the
corresponding reduced density matrices are further from thermal.
Nevertheless, the exponential scaling of the bond dimension with the variance also seems to hold in this case.

Finally, notice that the result about the scaling of the variance holds independent of the dimensionality of the problem.
Furthermore, since the bound on the rate of entanglement generation by a local Hamiltonian is general as well,
we expect that the bound of the entropy can also be generalized to higher dimensional systems.

\acknowledgments
We are thankful to A. Dymarsky for discussions.
This work was partly supported by the Deutsche Forschungsgemeinschaft (DFG, German Research Foundation) under Germany's Excellence Strategy -- EXC-2111 -- 390814868, and by the European Union through the ERC grant QUENOCOBA, ERC-2016-ADG (Grant no. 742102). 
M.C.B. acknowledges the hospitality of KITP, where part of this work was developed, and support from the National Science Foundation under Grant No. NSF PHY-1748958.  D.A.H. was supported in part by (USA) DOE grant DE-SC0016244.
\bibliography{MPSVariance}

\appendix

\end{document}